\documentclass[reprint,iopams,amsmath,amsfonts,amsgen,amssymb,aps]{iopart}
\iopamstrue 
\expandafter\let\csname equation*\endcsname\relax
\expandafter\let\csname endequation*\endcsname\relax

\usepackage{mathtools}
\usepackage{graphicx}
\usepackage{float}
\usepackage[labelfont=bf]{caption}
\usepackage{bm}
\usepackage{xcolor}
\usepackage[normalem]{ulem}
\usepackage{braket}
\usepackage{cite}
\usepackage{booktabs} 
\usepackage{hyperref}
\usepackage{amssymb}

\begin{document}
\title{A quantum model of lasing without inversion}

\author{Nicholas Werren\textsuperscript{1}, Erik M. Gauger\textsuperscript{1}, Peter Kirton\textsuperscript{2}}

\address{
 \textsuperscript{1}SUPA, Institute of Photonics and Quantum Sciences, Heriot-Watt University, Edinburgh, EH14 4AS, United Kingdom}
\address{\textsuperscript{2}Department of Physics and SUPA, University of Strathclyde, Glasgow, G4 0NG, United Kingdom
}
\ead{n.werren@hw.ac.uk}

\date{\today}

\begin{abstract}
Starting from a quantum description of multiple $\Lambda$-type 3-level atoms driven with a coherent microwave field and incoherent optical pumping, we derive a microscopic model of lasing from which we move towards a consistent macroscopic picture.
Our analysis applies across the range of system sizes from nanolasers to the thermodynamic limit of conventional lasing.
We explore the necessary conditions to achieve lasing without inversion in certain regimes by calculating the non-equilibrium steady state solutions of the model at, and between, its microscopic and macroscopic limits.
For the macroscopic picture, we use mean-field theory to present a thorough analysis of the lasing phase transition.
In the microscopic case, we exploit the underlying permutation symmetry of the density matrix to calculate exact solutions for $N$ 3-level systems.
This allows us to show that the steady state solutions approach the thermodynamic limit as $N$ increases, restoring the sharp non-equilibrium phase transition in this limit.
We demonstrate how the lasing phase transition and degree of population inversion can be adjusted by simply varying the phase of the coherent driving field.
The high level of quantum control presented by this microscopic model and the framework outlined here have applications to further understanding and developing nanophotonic technology.
\end{abstract}

\maketitle

\section{Introduction}

The textbook quantum description of a laser involves a gain medium of an ensemble of atomic 2-level systems (2LS) coupled to a single mode photonic cavity~\cite{Haken70, ScullyBook}.
The population inversion of the atomic ensemble required for lasing is achieved via an incoherent driving process. The microscopic origin of this inversion is usually via coherent pumping of transitions to other levels which, when adiabatically eliminated, give rise to the standard two-level model.
However, population inversion is not a necessary condition for lasing and lasing without inversion can be achieved through a degree of coherent control over 3-level systems~\cite{Scully89,Scully92,Zhu92}.

The various configurations available to 3-level atoms allow a variety of other non-trivial optical phenomena such as
coherent population trapping~\cite{Gray78, Xu08}, stimulated stimulated raman adiabatic passage~\cite{Gaubatz90, Kumar16}, and electromagnetically induced transparency~\cite{Boller91, Naeini11}.
However, when compared to 2-level systems, the many-body physics of 3-level atomic systems is relatively unexplored.
Such theory is not only interesting but fundamental to the development of future quantum technologies:
For instance, atomic clocks~\cite{Vanier05, Santra05, Zanon14}, the next generation of which incorporate lasers with many-body systems in order to engineer clock schemes with quantum noise-limited fractional stability~\cite{Pedrozo20}.
Beyond this, 3-level systems have proven to be a powerful tool in generating quantum memory for photons~\cite{Fleischhauer02, Lvovsky09}, atomic cooling~\cite{Aspect88, Marzoli94, Morigi00}, and quantum computing~\cite{Zhou02, Rao14, Higgins17}.

Of particular relevance to this work is nanoscale laser~\cite{Azzam20} and maser~\cite{Breeze18} technology.
As a consequence of their size, nanolasers exhibit a number of unique physical properties the most obvious of which is the low power consumption that comes as a result of the small gain medium~\cite{Ning19, Protsenko05} as well as the thresholdless operation that comes with being far from the thermodynamic limit~\cite{Rice94, Nomura09, Ojambati21}.
Such devices can achieve single-photon emission and thereby allow for the creation of photonic qubits; the encoding of quantum information in the photon state~\cite{Slussarenko19}. There are also possibilities to observe strong light-matter coupling effects~\cite{Leymann15, Andre2019} and squeezing~\cite{Mork20}. 
The physics of these lasing systems, from conventional laser to nanolaser, from macroscopic to microscopic is a vibrant and diverse area of research with a clear range of applications.
In this paper we explore the lasing behaviour across these regimes by modelling the inversionless lasing of a cavity QED model of 3-level atoms from the few- to the many-body limit.

The lasing phase transition of 3-level systems became the subject of intense inquiry following the discovery of lasing without population inversion (LWI)~\cite{Javan57, Mompart00, Richter20}.
Initial research into this area focused on gain increases created by utilising the recoil frequency shift experienced by an atom following the emission of a photon~\cite{Marcuse63}.
In two-level systems, this mechanism allows the amplification of radiation to occur without a larger population in the upper level than the lower level i.e.~without population inversion~\cite{Holt77}.

To identify population inversion in multi-level systems it is usual to group sets of levels into the ground and excited manifolds of an effective two-level system.
For instance, $\Lambda$-type 3-level systems are structured with a single upper energy level and two lower energy levels that are closely spaced (non-degenerate) or identical (degenerate).
In this case population inversion is defined as the collective population of the excited states being higher than that of the lower.
Moving beyond the recoil mechanism, research into LWI in the context of 3- and 4-level systems was significantly developed by Scully, Zhu, and collaborators in a series of seminal works \cite{Scully89,Scully92,Zhu92,Fearn92}.
Two distinct approaches to producing LWI in $\Lambda$-type 3-level systems were proposed:
The first utilises coherent population trapping (CPT) in degenerate and non-degenerate systems.
The system is prepared with initial quantum coherence between the two lower levels, which causes quantum interference effects that cancel absorption.
Stimulated emission increasingly becomes the only pathway available for the system and therefore results in an increased gain.
When the transitions from lower to upper state are completely excluded then any population in the higher energy level leads to lasing.
Thus, lasing is achieved without population inversion~\cite{Kocharovskaya92}.

In the second approach the two lower levels of a $\Lambda$-system are coupled by a coherent field~\cite{Fearn92}.
This set-up is known as a \textit{quantum beat-laser}~\cite{Scully87}.
This early model considers a beam of these atoms passing through a cavity field.
The coherent field splits and mixes the two atomic sublevels, inducing an interaction of the cavity with two resulting dressed states.
The coherences generated between the dressed states and the upper level produce similar interference effects to those observed in the case of CPT, suppressing absorption.

Following its theoretical conception, LWI of $\Lambda$-type 3-level systems has been experimentally investigated and subsequently observed in the case of CPT~\cite{Zibrov95, Peters96} and the quantum beat laser~\cite{Fearn92}.
Furthermore, recent work has shown how multiple $V$-type 3-level systems can generate many-body dark and nearly-dark states through dissipative stabilization of excited states~\cite{Lin21}.
Beyond this, recent studies have identified additional second-order phase transitions of 3-level systems separate to the laser threshold, revealing their relationships with the fundamental $U(1)$ symmetry~\cite{Minganti21}.
There is also interest in LWI of non-Hermitian systems, which can occur without any initial coherence~\cite{Doronin19, Miri19}.

As the number of components of a system increases, the dimension of the objects, such as the Liouvillian, that describes the  many-body physics grows exponentially. 
This makes calculating the dynamics and steady states challenging as large dimensions result in long computation times and impractical memory usage.
By using the permutation symmetry of the density matrix the size of the Liouvillian can be reduced~\cite{Richter15, Kirton17, Shammah18, Kirton19}.
This approach effectively increases the number of atoms for which a solution can be calculated.
As their number increases the dynamics of the atomic systems (and cavity) approach the thermodynamic limit wherein the lasing phase transition becomes increasingly sharp. Using this method allows us to examine the properties of the system at intermediate system sizes accessing this crossover region.

The structure of this paper is as follows:
In Sec.~\ref{sec:Mod}, we construct a microscopic model for a $\Lambda$-type 3-level system, including a semi-classical microwave drive and all of the necessary incoherent processes.
In Sec.~\ref{sec:MFSA}, we examine the behaviour in the thermodynamic limit by analysing the mean-field equations.
Linear stability analysis of these equations gives us access to the phase diagram.
This allows to calculate the location of the phase transition as a function of the phase of the microwave drive revealing the necessary parameters for LWI.
Sec.~\ref{sec:CB} moves beyond the mean-field equations by deriving the second order cumulant equations.
These give $1/N$ corrections to the mean-field equations which we are able to compare these results with exact solutions calculated using permutation symmetric methods.
This demonstrates how the behaviour of these systems approaches the thermodynamic limit.
Finally, in Sec.~\ref{sec:conc}, we present our conclusions.

\section{Model} \label{sec:Mod}

\begin{figure}
\captionsetup{width=.9\linewidth}
  \lineskip=-\fboxrule
  \fbox{\begin{minipage}{\dimexpr \textwidth-2\fboxsep-2\fboxrule}
    \centering
    \includegraphics[height=3.6 cm, angle=0]{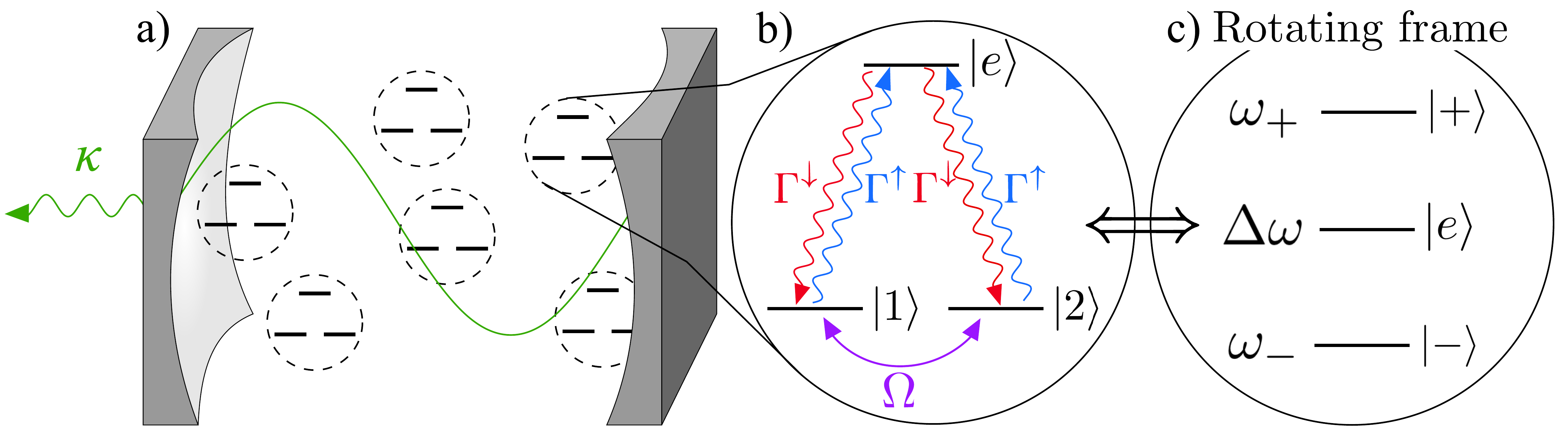}
    \abovecaptionskip=0pt
    \caption{Schematic diagram of the model used. Panels~a) and b) show the decay and drive processes of several $\Lambda$-type 3-level systems in a cavity with relaxation rate $\kappa$. The upper energy level $\ket{e}$ of the atomic systems are incoherently pumped at rate $\Gamma^{\uparrow}$ and decay at rate $\Gamma^{\downarrow}$. The two lower atomic levels, $\ket{1}$ and $\ket{2}$, are coupled by a coherent microwave field with strength $\Omega$. Panel~(c) depicts the 3-level atoms in the microwave-dressed eigenbasis of the Hamiltonian and in the rotating frame of the cavity transition. Note the incoherent processes are not shown in this panel.
    }
    \label{fig: 3LSdriven}
  \end{minipage}}
\end{figure}

We begin our analysis by constructing a microscopic model, as illustrated in Figure~\ref{fig: 3LSdriven}a), that is able to capture the physics of a quantum beat laser~\cite{Fearn92}.
To this end, we consider a collection of degenerate $\Lambda$-type atomic 3-level systems {where the transition between ground states is coherently driven by an applied microwave field} which we treat semi-classically.
The upper level and the two ground levels of each `atom' are labelled as $\ket{e}$, $\ket{1}$, and $\ket{2}$ respectively.

The Hamiltonian for $N$ such systems interacting with a single cavity mode is given by:
\begin{equation} \label{eq:OHam}
    H = H_F + \sum_{i}^{N}H_{AF}^{(i)}  + H_{M}^{(i)} ,
\end{equation}
where $H_F$ is the field Hamiltonian and $H_{AF}^{(i)}$ describes the $i$th atom and its interaction with the cavity and $H_{M}^{(i)}$ is the interaction with the semi-classical microwave drive.
These are given by:
\begin{gather}
    H_{F} =  \nu a^{\dag} a, \\
\label{eg:HAF1}
    H_{AF}^{(i)} = \omega_e \rho_{ee}^{(i)}+\frac{g}{\sqrt{N}}\left[ a (\rho_{e1}^{(i)} + \rho_{e2}^{(i)} )  + \text{H.c.} \right],\\
    H_{M}^{(i)} = \Omega \left( \text{e}^{i\phi}\rho_{21}^{(i)} + \text{H.c.} \right).
\end{gather}
The cavity is described by the annihilation operator $a$ and has frequency $\nu$.
To describe the atomic degrees of freedom we use the notation $\rho^{(i)}_{ab} = \ket{a}\bra{b}_i$ where each state $\ket{k}$ has an associated energy $\omega_{k}$ ($\hbar=1$).
For simplicity we assume that the two lower atomic energy levels are degenerate and use this as our reference energy such that $\omega_{1}= \omega_{2}=0$.
Furthermore, we assume that the coupling strength, $g$, between the light and matter is weak compared to the bare cavity and atomic transition frequencies and so we perform the rotating-wave approximation on this term, ignoring the counter-rotating terms\cite{ScullyBook}.
The factor of $\sqrt{N}$ in the light matter coupling term ensures a well-defined thermodynamic limit when taking $N\to\infty$. With this scaling when the system size is changed the collective light mater coupling remains constant as the number of emitters is increased\cite{Wang73, Emary03, Kirton19}.
Following the approach from Ref.~\cite{Scully89}, we have already moved to a rotating frame and performed a rotating wave approximation\footnote{Note that this removes fast oscillating terms in the Hamiltonian not only between the ground state levels but also from the light-matter coupling terms.} with respect to the microwave field which is described by the Rabi frequency $\Omega$ which has a phase of $\phi$ relative to the light-matter coupling.
A gauge transformation which removes the phase from this term in the Hamiltonian would introduce an equivalent phase on the light-matter coupling $g$.
We note that the model outlined here does not include any microwave induced relaxation between the two lower levels.

To include the effects of incoherent processes along with the coherent ones included in the above Hamiltonian we write a Lindblad form quantum master equation for the density matrix $\rho$ of the combined cavity and atomic degrees of freedom.
This takes the form~\cite{Scully92, Wiseman09}:
\begin{multline} \label{eq:QME1}
    \dot \rho = -i \left [H, \rho \right] + \kappa \mathcal{D}[a]\rho + \Gamma^{\uparrow}\sum^{N}_{i}  \left(\mathcal{D}[\rho_{e1}^{(i)}]+\mathcal{D}[\rho_{e2}^{(i)}]\right)\rho  \\
    + \Gamma^{\downarrow}\sum^{N}_{i} \left(\mathcal{D}[\rho_{1e}^{(i)}]+\mathcal{D}[\rho_{2e}^{(i)}]\right)\rho,
\end{multline}
where the usual Lindblad superoperator is $\mathcal{D}[x]\rho=x\rho x^\dagger-\frac{1}{2}\{x^\dagger x,\rho\}$.
We have introduced cavity relaxation at rate $\kappa$, and a common incoherent pumping rate $\Gamma^{\uparrow}$ from both lower levels to $\ket{e}$, mirrored by an incoherent loss of atomic excitations at rate $\Gamma^{\downarrow}$. These incoherent processes can be realised by using a coherent drive to an additional level, as is standard for regular 2-level lasing~\cite{ScullyBook}, or as the result of a weak coherent drive applied directly to the relevant transition. The second case would, however, restrict the range of parameters to $\Gamma^{\uparrow}\leq\Gamma^{\downarrow}$.

This model has a $U(1)$ symmetry which corresponds to shifting the phase of the cavity and atomic coherences $a\to \text{e}^{i\theta}a$, $\rho_{e1}\to \text{e}^{-i\theta}\rho_{e1}$, and $\rho_{e2}\to \text{e}^{-i\theta}\rho_{e2}$.  The lasing transition of this model occurs when this symmetry is spontaneously broken.
For the purpose of comparison a parallel description for the case of 2-level instead of 3-level systems is provided in \ref{app:2QM}.

This model could be realised in a variety of physical systems. An obvious approach utilises cold atoms trapped in optical cavities. In these systems fine control over the required level structure, interaction strengths and decay rates can be achieved~\cite{Gothe2019}. Other potential platforms where this kind of physics could be realised includes atomic ensembles in vapour cells~\cite{Thomas2019} or rare earth ion doped solid-state ensembles~\cite{Kutluer2019} where $\Lambda$-like Raman level structures have been shown to enable quantum memories. Impurities in diamond~\cite{Breeze18} also hold promise if a suitable level structure can be engineered.

\section{Mean-Field Theory} \label{sec:MFSA}

In the thermodynamic limit of a large ensemble of emitters,  $N\to\infty$, it is well known that this class of model is described exactly by mean-field theory~\cite{Haken70}.
This allows us to characterise the point at which the lasing phase transition occurs as a function of the parameters of the master equation.
In this limit the transition is identified by the point at which the cavity population per atom $\braket{a^{\dag}a}/N$ becomes finite; below the lasing transition, in the normal state, $\braket{a^{\dag}a}\sim \sqrt{N}$ while above the transition $\braket{a^{\dag}a}\sim N$.
To derive the mean-field equations we make use of the fact that all of the emitters are identical and then use the master equation, \eqref{eq:QME1} to find the evolution of each of the required operators using the relation $\braket{\dot x}=\Tr[x \dot \rho]$.
In the limit where mean-field theory is valid we may neglect second order and higher cumulants and write a closed set of equations by breaking second order correlations such as $\braket{a \rho_{e1}}=\braket{a} \braket{ \rho_{e1}}$\cite{Haken70, Kirton19}.
We will return to the effect that approximation has at finite $N$ in Sec.~\ref{sec:CB}.
We hence obtain equations that describe the evolution of the expectation values of the complex photon amplitude and atomic degrees of freedom:
\begin{equation}\label{eq: MFE_rhoe1}
    \braket{\dot\rho_{e1}} = i \left[ \omega_{e}\braket{\rho_{e1}} + \frac{g}{\sqrt{N}}\braket{a^{\dag}}\left(\braket{\rho_{11}} -  \braket{\rho_{ee}} + \braket{\rho_{21}}\right) 
    - \Omega \text{e}^{-i\phi}\braket{\rho_{e2}}\right] - \Gamma_\phi \braket{\rho_{e1}}, 
\end{equation} 
\begin{equation} \label{eq: MFE_rhoe2}
   \braket{\dot\rho_{e2}} = i  \left[\omega_{e}\braket{\rho_{e2}} + \frac{g}{\sqrt{N}}\braket{a^{\dag}}\left(\braket{\rho_{22}} - \braket{\rho_{ee}} + \braket{\rho_{12}} \right) 
    - \Omega \text{e}^{i\phi}\braket{\rho_{e1}}\right] - \Gamma_\phi \braket{\rho_{e2}}, 
\end{equation}
\begin{equation} \label{eq: MFE_rho12}
    \braket{\dot\rho_{12}} = i\left[- \frac{g}{\sqrt{N}}\left( \braket{a^{\dag}}\braket{\rho_{1e}} -  \braket{a}\braket{\rho_{e2}}\right) + \Omega \text{e}^{i\phi}\left(\braket{\rho_{22}} - \braket{\rho_{11}} \right)\right] - \Gamma^{\uparrow}\braket{\rho_{12}},
\end{equation}
\begin{equation} \label{eq: MFE_rhoee}
    \braket{\dot\rho_{ee}} = -\frac{g}{\sqrt{N}}\left[ i\braket{a}(\braket{\rho_{e1}} + \braket{\rho_{e2}}) + \text{c.c.} \right]
    + \Gamma^{\uparrow} \left(\braket{\rho_{11}} + \braket{\rho_{22}} \right) - 2\Gamma^{\downarrow} \braket{\rho_{ee}},
\end{equation}
\begin{equation} \label{eq: MFE_rho11}
    \braket{\dot\rho_{11}} = \left[ i\left(\frac{g}{\sqrt{N}}\braket{a}\braket{\rho_{e1}} + \Omega \text{e}^{i\phi}\braket{\rho_{21}}\right) + \text{c.c.} \right] - \Gamma^{\uparrow} \braket{\rho_{11}} + \Gamma^{\downarrow} \braket{\rho_{ee}},
\end{equation}
\begin{equation} \label{eq: MFE_rho22}
    \braket{\dot\rho_{22}} = \left[i\left( \frac{g}{\sqrt{N}}\braket{a}\braket{\rho_{e2}} - \Omega \text{e}^{i\phi}\braket{\rho_{21}} \right)  + \text{c.c.}\right] - \Gamma^{\uparrow} \braket{\rho_{22}} + \Gamma^{\downarrow} \braket{\rho_{ee}},
\end{equation}
\begin{equation} \label{eq: MFE_a}
    \braket{\dot{a}} = -i\left[ \nu \braket{a} + g\sqrt{N}\left( \braket{\rho_{1e}} + \braket{\rho_{2e}}\right) \right] - \frac{\kappa}{2}\braket{a},
\end{equation}
where $\Gamma_\phi = (\Gamma^{\uparrow} + 2\Gamma^{\downarrow})/2$ is the total atomic dephasing rate.
This is induced by the presence of the two incoherent rates, $\Gamma^{\uparrow}$ and $\Gamma^{\downarrow}$, which describing energy relaxation processes and hence give rise to an associated dephasing process.
From these equations we can extract the conditions under which the lasing phase transition occurs.
To derive an expression for the location of the phase transition we consider the linear stability of the normal state fixed point.
It is straightforward to show that the coherences $\braket{a}_{ns}= 0$, $\braket{\rho_{1e}}_{ns} = 0$, $\braket{\rho_{2e}}_{ns} = 0$ and $\braket{\rho_{12}}_{ns} = 0$ and populations $\braket{\rho_{ee}}_{ns} = \Gamma^{\uparrow} /{2\Gamma_\phi}$, $\braket{\rho_{11}}_{ns} = \braket{\rho_{22}}_{ns} ={\Gamma^{\downarrow} }/ {2\Gamma_\phi}$ are always a solution of Equations~\eqref{eq: MFE_rhoe1}-\eqref{eq: MFE_a}.
We can then linearise around this fixed point to find the point at which it becomes unstable and hence the system enters the lasing phase.
We find that the only terms which give non-trivial contributions to this linear stability are $\delta a = \braket{a} - \braket{a}_{ns}$, $\delta \rho_{1e} = \braket{\rho_{1e}} - \braket{\rho_{1e}}_{ns}$ and $\delta \rho_{2e} = \braket{\rho_{2e}} - \braket{\rho_{2e}}_{ns}$~\footnote{All other terms are either uncoupled complex conjugates of these or never contribute to the instability.}.
Hence we can write the linearised equation for these fluctuations
\begin{equation}
    \frac{d}{dt} \begin{pmatrix}
\delta a\\
\delta \rho_{1e}\\
\delta \rho_{2e}\\
\end{pmatrix} = J \begin{pmatrix}
\delta a\\
\delta \rho_{1e}\\
\delta \rho_{2e}\\
\end{pmatrix},
\end{equation}
where the Jacobian matrix is
\begin{equation}\label{eq:Jacobian1}
J
=
\begin{pmatrix}
-i \nu - \frac{\kappa}{2} & -ig\sqrt{N} & -ig\sqrt{N} \\
ig \left( \braket{\Delta_{e1}}_{ns} - \braket{\rho_{12}}_{ns}\right)/\sqrt{N} & i\omega_{e} - \Gamma_\phi & i \Omega \text{e}^{i \phi }  \\
ig \left( \braket{ \Delta_{e2}}_{ns}- \braket{\rho_{21}}_{ns}\right)/\sqrt{N}  & i \Omega \text{e}^{-i \phi }  & i\omega_{e} - \Gamma_\phi \\
\end{pmatrix},
\end{equation}
and we have introduced the operators
\begin{align}
    \Delta_{e1} = \rho_{ee} - \rho_{11}, && \Delta_{e2} = \rho_{ee} - \rho_{22}.
\end{align}
When the real part of one of the eigenvalues of Equation~\eqref{eq:Jacobian1} becomes positive the normal state is unstable and the cavity is occupied, thereby allowing us to identify the phase transition from the eigenvalues of the characteristic polynomial of the Jacobian:
\begin{multline} \label{eq:cubic3LS1}
\lambda^{3} + \left( 2\Gamma_{\phi} + i(\nu + 2\omega_e) + \frac{\kappa}{2} \right)  \lambda^{2} \\
+  \left[ g^{2} \frac{(\Gamma^{\downarrow} - \Gamma^{\uparrow})}{\Gamma_{\phi}} + (\Gamma_{\phi} + i\omega_{e}) \left(\Gamma_{\phi} + i(2\nu + \omega_{e}) + \kappa \right)  + \Omega^{2}\right] \lambda \\
+ g^{2} \frac{(\Gamma^{\downarrow} - \Gamma^{\uparrow})}{\Gamma_{\phi}}\left[\Gamma_{\phi} + i\left(\omega_{e} +  \Omega \cos\left(\phi\right)\right) \right] + \frac{1}{2} \left(\kappa + 2i\nu \right)\left[\left(\Gamma_{\phi} + i\omega_{e}\right)^{2} + \Omega^{2} \right] =0,
\end{multline}
where $\lambda$ is an eigenvalue of $J$.
Therefore, solving for $\lambda$ corresponds to finding the roots of Equation (\ref{eq:cubic3LS1}).
The location of the phase transition as a function of the pumping rate $\Gamma^\uparrow$ has a non-trivial dependence on the phase of the applied microwave drive, $\phi$.
Hence, the structure of the eigenvalues is more complicated than that of a 2-level system which we present for comparison in \ref{app:2QM}.

\subsection{Mean-Field Results}

We can now use the mean-field equations and linear stability analysis to examine the optimal conditions for observing lasing in this model.
To develop intuition for the underlying physics it is useful to first look in more detail at the structure of the Hamiltonian.
The splitting of the energy levels by the microwave field effectively shifts the resonant cavity frequency.
This has a significant impact on the lasing phase transition.
In order to capture this behaviour and identify the shifted resonant frequency, we rewrite the Hamiltonian, Equation~\eqref{eq:OHam}, in terms of the following dressed states which diagonalises the terms which do not involve the cavity
\begin{equation}
    \ket{+} =  \frac{1}{\sqrt{2}}\left(\text{e}^{i\phi/2} \ket{2} + \text{e}^{-i\phi/2} \ket{1} \right)
\end{equation}
and
\begin{equation}
    \ket{-} =  \frac{1}{\sqrt{2}}\left(\text{e}^{i\phi/2} \ket{2} - \text{e}^{-i\phi/2} \ket{1}\right).
\end{equation}
Using these states we move to a frame rotating at the cavity frequency to give the Hamiltonian
\begin{multline} \label{eq:dressedhamf}
    H_{R} = \sum_{i}^{N} \Delta \omega \rho_{ee}^{(i)}+ \omega_{+}\rho_{++}^{(i)} + \omega_{-}\rho_{--}^{(i)} \\
    +  \frac{g}{\sqrt{N}}\left( a \left[\cos \left(\frac{\phi}{2} \right)\rho_{e+}^{(i)} - i\sin \left(\frac{\phi}{2} \right)\rho_{e-}^{(i)}\right] + \text{H.c.} \right)
\end{multline}
where $\Delta\omega = \omega_{e} - \nu$ and $\omega_{\pm} = \pm \Omega$.
This is illustrated in Figure~\ref{fig: 3LSdriven}(c).

\begin{figure}
\captionsetup{width=.9\linewidth}
  \lineskip=-\fboxrule
  \fbox{\begin{minipage}{\dimexpr \textwidth-2\fboxsep-2\fboxrule}
    \centering
    \includegraphics[height=4.5 cm, angle=0]{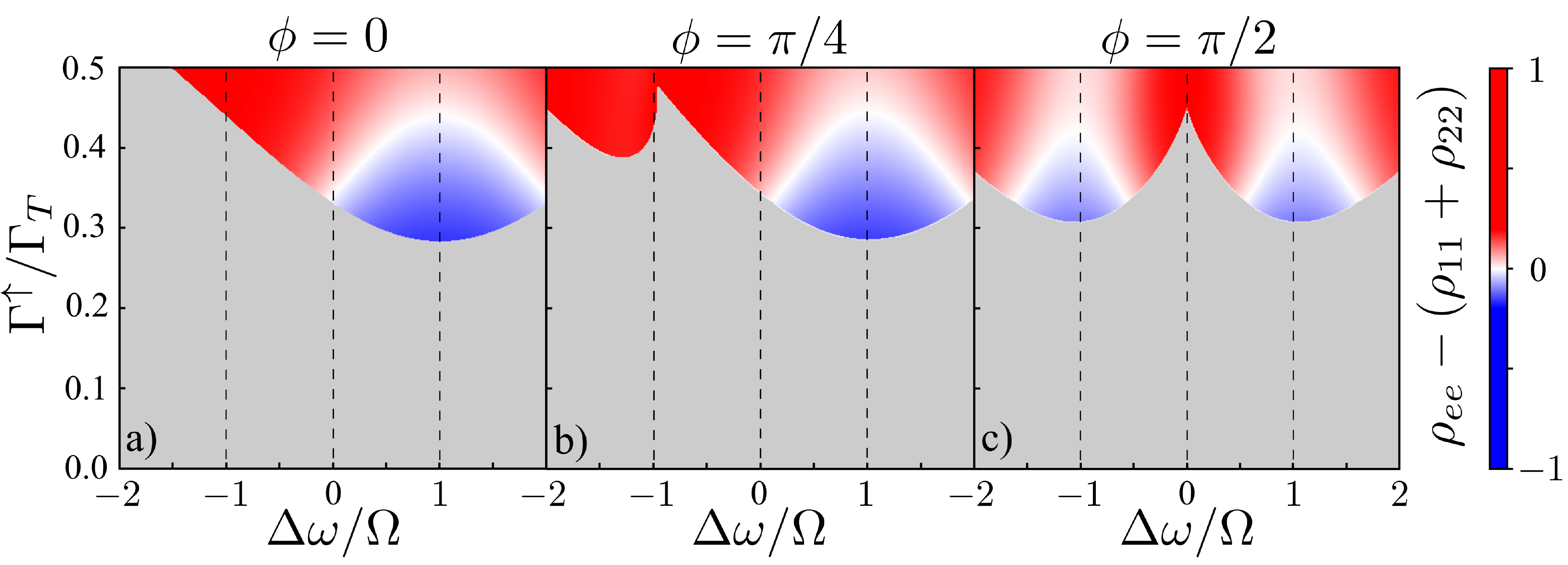}
    \abovecaptionskip=0pt
    \caption{Mean-field phase diagram showing the lasing regions at different values of $\phi$. In the white region the normal state is stable, while in the coloured regions a lasing instability occurs and the Jacobian matrix has a pair of eigenvalues with positive real part. In this region the degree of inversion, and therefore lasing without inversion, is indicated by the colour gradient. Excluding those stated above, the parameters are $g=0.9\Omega$, $\kappa=0.8\Omega$, and $\Gamma_T = \Omega$.}
    \label{fig:PhiPhaseScan}
  \end{minipage}}
\end{figure}

Outside of certain parameter regimes the characteristic polynomial, Equation~\eqref{eq:cubic3LS1}, does not have straightforward solutions and so we numerically compute the eigenvalues of the Jacobian, Equation~\eqref{eq:Jacobian1}, to find the phase diagram as a function of the atomic pumping rate $\Gamma^\uparrow$ and atom-cavity detuning $\Delta\omega$.
It is important to note that both $\Gamma^\uparrow$ and $\Gamma^\downarrow$ describe two individual incoherent processes, as such we define the total rate $\Gamma_{T} = 2(\Gamma^\uparrow + \Gamma^\downarrow)$ which allows us to normalise the rates.
Examining how this phase diagram changes for different values of the phase of the microwave drive $\phi$ provides a complete picture of the lasing behaviour in the thermodynamic limit.
Figure~\ref{fig:PhiPhaseScan} shows these phase diagrams for different values of $\phi$.
The $y$-axis is in terms of $\Gamma^{\uparrow}/\Gamma_{T}$.
We vary this quantity while keeping $\Gamma_T$ constant such that when $\Gamma^{\uparrow}/\Gamma_{T}=0$ there is no pumping and when $2\Gamma^{\uparrow}/\Gamma_{T}=0.5$ there is no atomic loss.
In these diagrams we highlight where $\Delta \omega = \pm \Omega$ since this corresponds to the detunings where the cavity is on resonance with the microwave-dressed atomic states described above.

In Figure~\ref{fig:PhiPhaseScan}(a), for $\phi=0$ we see a single minimum in the lasing region at $\Delta \omega=\Omega$.
This can be understood as follows:
As previously stated, the coherent microwave field mixes the lower two atomic energy levels with driving strength $\Omega$. This results in the hierarchy of energy levels shown in Figure~\ref{fig: 3LSdriven}b) and c) where $\omega_{\pm}=\pm\Omega$.
As such this forces a particular phase dependence between the state of the atoms and the emitted photons.
The term describing the interaction between the cavity and the atomic transitions in Hamiltonian~\eqref{eg:HAF1} accounts for this dependence and thereby defines a phase reference for the microwave field.
As a result, applying the microwave drive with phase $\phi = 0$ favours the formation of the $\ket{+}$ state and hence we observe that lasing occurs when the cavity is tuned to the frequency of this state.
In the opposite limit of $\phi=\pi/2$, shown in Figure~\ref{fig:PhiPhaseScan}(c), we see two symmetric local minima at $\Delta\omega=\pm\Omega$ (with some corrections proportional to the cavity relaxation rate $\kappa$).
In that case the interplay between light-matter interaction and the microwave drive favours the formation of an equal superposition of the $\ket{+}$ and $\ket{-}$ states and so lasing occurs at both frequencies.
When the microwave phase is between these limits as in Figure~\ref{fig:PhiPhaseScan}(b), where $\phi=\pi/4$, we see a smooth crossover between these limits.
In this case the minima at $\Delta\omega=\Omega$ remains largely unchanged.
In all cases we see that close to the minimum required pumping the lasing state is stable while the inversion is still negative. As the pump power required is increased the observed inversion level also increases.
The results shown here outline the range of possible behaviours, for completeness however, in \ref{app:3MFA} we also show how the phase diagram depends on the total rate $\Gamma_T$ and the microwave phase $\phi$.

\begin{figure}
\captionsetup{width=.9\linewidth}
  \lineskip=-\fboxrule
  \fbox{\begin{minipage}{\dimexpr \textwidth-2\fboxsep-2\fboxrule}
    \centering
    \includegraphics[height=5.0 cm, angle=0]{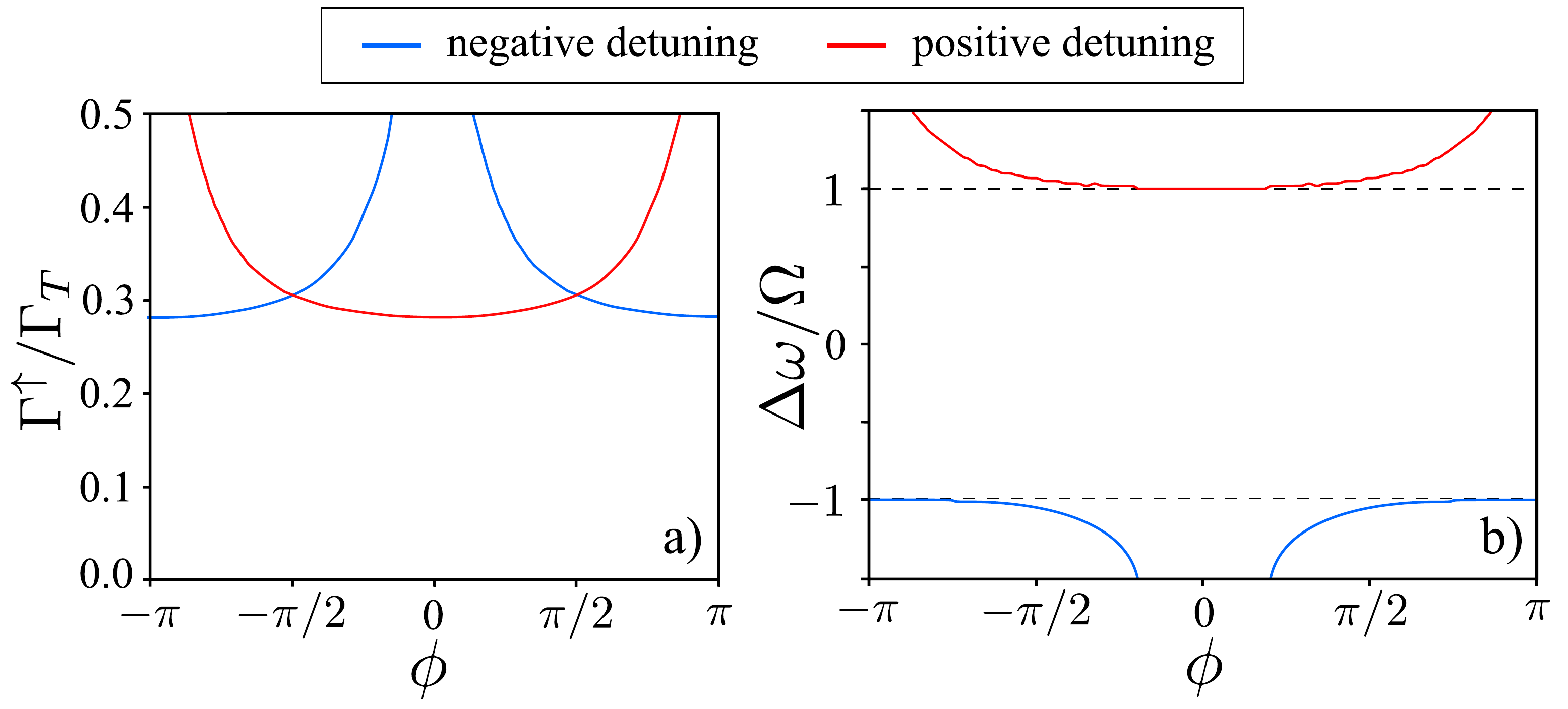}
    \abovecaptionskip=0pt
    \caption{Location of the minimum pumping strength required to reach the lasing phase transition. The red lines show the location of the minimum at positive detuning, the blue show the equivalent  result for negative detuning. Panel a) gives the pump power of the minima while panel b) shows the frequency. All other parameters are the same as in Figure~\ref{fig:PhiPhaseScan}.}
    \label{fig:MinimaTracking}
  \end{minipage}}
\end{figure}

To further understand how the location of these minima behave we track their locations in terms of both their frequency and pump strength as a function of the microwave phase $\phi$.
We capture this behaviour in Figure~\ref{fig:MinimaTracking} where the location of the minimum at positive detuning is shown in red while the one at negative detuning is in blue.
We see that there are large plateaus close to $\phi=0$ and $\phi=\pm\pi$ where the microwave drive allows lasing from only one of the dressed states.
This results in the lowest value of the incoherent pump strength at each of these points.
At the centre of these plateaus only a single positively or negatively detuned minimum exists, its associated $\Delta\omega$ corresponds to the cavity mode being exactly resonant with one of the atomic dressed states, as expected.
A shift away from that to larger detuning is observed in the overlap regions, and is particularly pronounced for the respective `higher lying' (i.e.~stronger pumping to reach the phase transition) minimum.
From these results we are able to find an exact description of the critical pumping required to achieve lasing for particular cases.
Exactly at points $\phi=0,\pm\pi$ we are able to find a simple analytic expression for the critical pumping rate from the characteristic polynomial in Equation~\eqref{eq:cubic3LS1}:
\begin{equation}
    \Gamma_{\text{crit}}^{\uparrow} = \Gamma_{T} + 8\frac{g^{2}}{\kappa}\left(1 -  \sqrt{1+\frac{3\Gamma_{T}}{16} \frac{ \kappa}{g^{2}}}\right),
\end{equation}
which is valid exactly on resonance.
Here, we are incoherently pumping from both lower levels such that in the limit $g \gg \kappa$ we find $\Gamma_{\text{crit}}^{\uparrow}/\Gamma_{T} = 1/4$.
For the same limit in the case of the 2-level system, where the incoherent pumping is only  from a single lower level, we find instead the equivalent result $\Gamma_{\text{crit}}^{\uparrow}/\Gamma_{T} = 1/2$.

\begin{figure}
\captionsetup{width=.9\linewidth}
  \lineskip=-\fboxrule
  \fbox{\begin{minipage}{\dimexpr \textwidth-2\fboxsep-2\fboxrule}
    \centering
    \includegraphics[height=9.0 cm, angle=0]{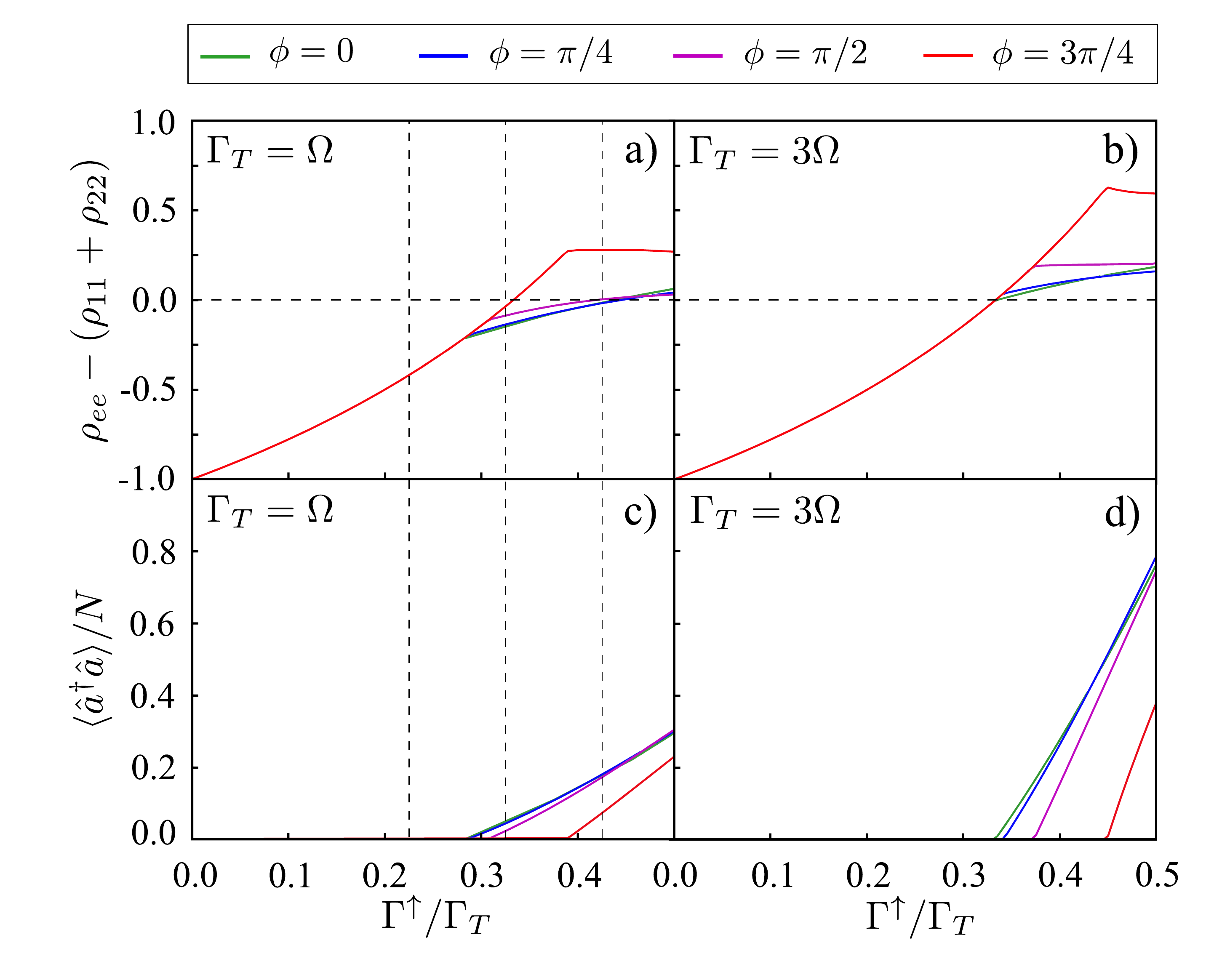}
    \abovecaptionskip=0pt
    \caption{Atomic population inversion and cavity population as a function of pumping strength for different values of the total atomic decoherence rate. The vertical dashed lines show the values of $\Gamma^{\uparrow}/\Gamma_{T}$ used in Figure~ \ref{fig:methodcomp1}. Upper row: Population inversion at the indicated values of $\phi$ and $\Gamma_T$. The horizontal dashed line shows the point at which the excited and ground state populations are equal. Lower row: Cavity population for the same parameters. Below the transition the phase of the microwave drive has no effect and thus multiple lines overlap. The detuning is fixed to the minimum pumping threshold described in Figure~\ref{fig:PhiPhaseScan}: For $\phi=0$ and $\phi=\pi/4$ this is $\Delta \omega = \Omega$. Cavity effects on the threshold become more apparent for $\phi=\pi/2$ and $\phi=3\pi/4$ where $\Delta \omega = 1.08 \Omega$ and $\Delta \omega = 1.28 \Omega$ respectively. Other parameters are $g= 0.9\Omega$ and $\kappa = 0.8\Omega$.}
    \label{fig:3LSAtomPop}
  \end{minipage}}
\end{figure}

To explore this further it useful to use the full mean-field equations~\eqref{eq: MFE_rhoe1}--\eqref{eq: MFE_a} to find the steady state population inversion and cavity population generated by this model as a function of incoherent pumping $\Gamma^{\uparrow}/\Gamma_{T}$.
We define the population inversion as the difference between the excited state population and the total population in the ground state manifold $\rho_{ee}- (\rho_{11}+\rho_{22})$.
In Figure~\ref{fig:3LSAtomPop} we show how this changes  as the phase of the microwave field is varied from $\phi = 0$ to $\phi=3\pi/4$ for two different values of the total atomic decoherence rate $\Gamma_T$.
The main result here is that we observe a strong dependence on the nature of the phase transition with changing $\phi$.
Lasing without inversion when the phase of the microwave drive is tuned to $\phi\lesssim \pi/2$ such that the $\ket{+}$  state is favoured.
This only occurs when $\Gamma_T$ is small enough since when this is increased a larger population is required in the excited state to overcome the decay.
When $\phi\gtrsim \pi/2$ significant inversion is required before the phase transition occurs.
This is because, for these values of $\phi$, the microwave drive populates the $\ket{-}$ state which is far detuned.

A significant difference from the usual 2-level laser model is the lack of mode-locking.
As we see in \ref{app:2QM} above the lasing transition the population inversion locks to a fixed value and all of the additional energy from the pumps goes into the cavity field.
We do observe behaviour close to mode-locking for $\phi=3\pi/4$ and $\Gamma_T=\Omega$, whereas for $\Gamma_T=3\Omega$ this occurs at $\phi=\pi/2$.
This effect originates from a different mechanism to the usual case.
In the case where the laser is far detuned (for the largest value of $\phi$ shown) we see that the population inversion \textit{decreases} above the transition threshold.
When there is a large coherent field inside the cavity the atomic dynamics becomes too fast for the microwave drive to have any effect and so the dressing of the atomic states is reduced and the inversion along with it.
In the limit of a very large occupation of the cavity the correct picture for the atomic system becomes the bare states.

\section{Finite $N$ Results} \label{sec:CB}

Whilst mean-field theory provides useful information regarding the phase transition {for traditional lasing where $N\to\infty$}, it is not appropriate for small {or intermediate} numbers of 3-level systems.
To analyse this limit in the following we derive a set of second-order cumulant equations which provide a systematic way of calculating corrections of order $1/N$ to the mean-field results allowing access to the intermediate $N$ regime.
Along with this we also provide exact numerical solutions to our microscopic model for {small} $N$.
These exact simulations make use of the permutation symmetry of the Liouvillian~\cite{Shammah18} allowing us to simulate around $N\sim10$ 3-level systems.

To derive the second-order cumulant equations we retain correlations between pairs of operators and instead break the third order moments into products of first and second order correlations~\cite{Kubo62, Kirton19, Plankensteiner22}.
This procedure then produces a closed set of equations for the first and second moments which in the limit $N\to\infty$ reproduce the mean-field equations and include $1/N$ corrections which vanish in this limit.
The full set of equations is cumbersome and so we give full details of the derivation in \ref{app:3MFA}.
Furthermore, a parallel derivation of the cumulant equations for the equivalent model with 2-level systems is provided in \ref{app:2QM} for comparison.
\begin{figure}
\captionsetup{width=.5\linewidth}
\centering
  \lineskip=-\fboxrule
  \fbox{\begin{minipage}[t]{0.7\linewidth}
    \centering
    \includegraphics[height=5.0 cm, angle=0]{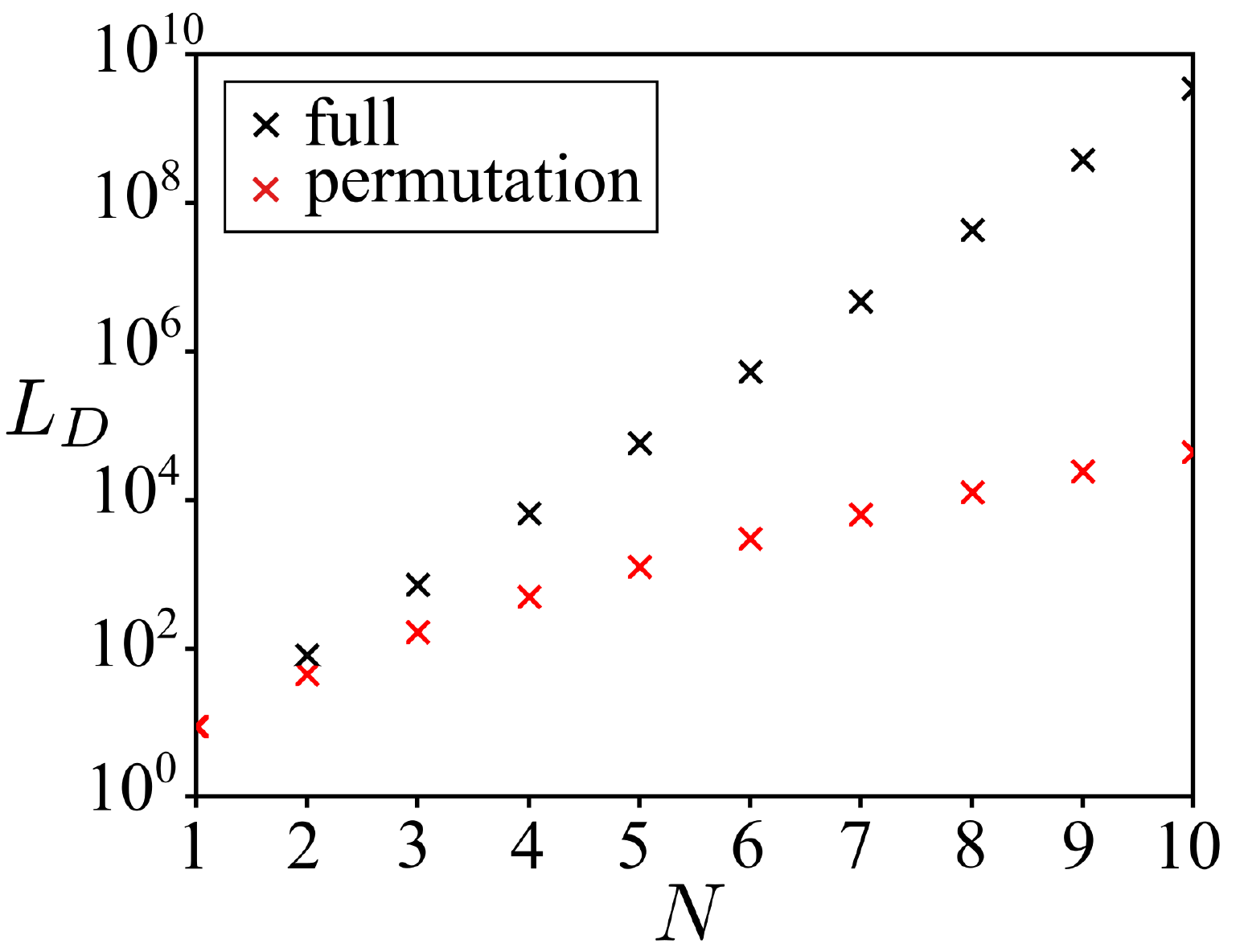}
    \abovecaptionskip=0pt
    \caption{The dimension of the Liouville space, $L_{D}$, prior to the inclusion of the {cavity F}ock space, constructed for $N$ 3-level systems by using permutation symmetry (red crosses) compared with the full Liouvillian (black crosses).}
    \label{fig:hilspace}
  \end{minipage}}
\end{figure}

For small values of $N$ we make use of exact simulations of the master equation~\eqref{eq:QME1} to find the steady state density operator.
Directly simulating this scales exponentially with the number of 3-level systems included and is difficult even for $N\sim5$.
Instead we make use of the permutation symmetry of the Liouvillian which arises because each atom is identical and couples identically to the cavity mode.
This approach is fully outlined in Refs.~\cite{Kirton17, Shammah18} whilst the code which our results are based on  can be found at~\cite{KirtonGit17}.
This technique has been previously applied to a variety of models with 2-level systems.
For example plasmons~\cite{Richter15}, spin ensembles~\cite{Chase08}, subradiant states~\cite{Damanet16, Gegg18}, superradiant phase transitions in the Dicke model~\cite{Kirton17, Kirton18} and fully connected Ising models~\cite{Wang21}.
The generalisation of these approaches to 3-level systems is  straightforward.
We show the compression achieved by using this method in Figure~\ref{fig:hilspace}.
The dimension of the Liouville space, $L_D$, that describes $N$ {copies of our} 3-level system scales exponentially as $3^{2N}$ however using the permutation symmetry this scaling is reduced to polynominal in the system size.
For example at $N=2$, the permutation approach produces reduces $L_D$ from 81 to 45 while at $N=8$, the full space has $L_D=43,046,721$ which is reduced down to a much more manageable $L_D=12,870$ by the permutation symmetry.
We note here that the full simulations need to include the Hilbert space of the cavity resulting in an increase to $L_D$ by a factor of $P^{2}$, where $P$ is the number of Fock states used to represent the cavity field.

\begin{figure}
\captionsetup{width=.9\linewidth}
  \lineskip=-\fboxrule
  \fbox{\begin{minipage}{\dimexpr \textwidth-2\fboxsep-2\fboxrule}
    \centering
    \includegraphics[height=18.0 cm, angle=0]{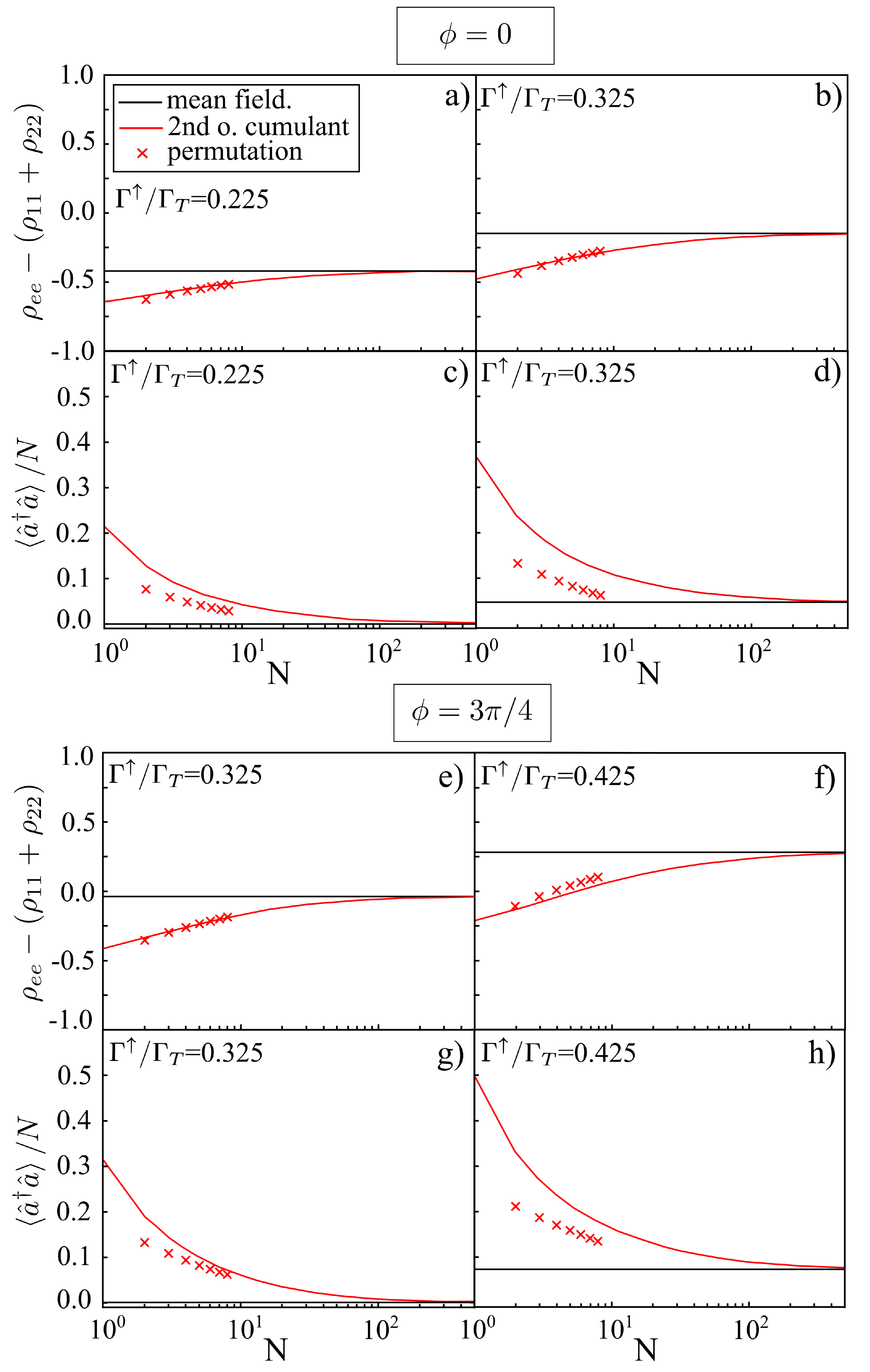}
    \abovecaptionskip=0pt
    \caption{Agreement between the second order cumulant and exact numerical (permutation) solutions with mean-field theory at different $\Gamma^{\uparrow}$, $\phi$, and $N$. We have set $\Gamma_{T}=\Omega$, and all other parameters are the same as in Figure~\ref{fig:3LSAtomPop}.}
    \label{fig:methodcomp1}
  \end{minipage}}
\end{figure}

We begin by looking at how the results at finite $N$ scale towards the thermodynamic limit.
In Figure~\ref{fig:methodcomp1} we show how the second-order cumulant results converge upon those of the mean-field as $N$ increases for a representative set of parameters both above and below the  lasing phase transition and for different values of $\phi$.
For each case we calculate the reduced photon number $\braket{a^{\dag}a}/N$ and population inversion.
Alongside these results we also show the exact numerical steady state solutions of the master equation~\eqref{eq:QME1} for $N$ systems, truncating the cavity Hilbert space to $P=10$ Fock states.
This allows for a direct comparison with the mean-field theory discussed in the previous section and captures the importance of cross-correlations to the microscopic description of this system.

For both $\phi = 0$ and $\phi = 3\pi/4$ we see very good agreement between the exact permutation solutions and the corresponding second-order cumulant solutions, even at relatively small values of $N$.
The agreement is generally better for the atomic inversion but even the results for the cavity photon number show convergence as $N$ increases.
At the same time we see that the cumulant results approach the mean-field limit as $N$ increases.
We thus show that by using a combination of these techniques it is possible to access the behaviour across the whole range of system sizes from the microscopic $N\sim1$ to the macroscopic $N\to\infty$.
We note however that the parameters chosen here are fairly specific.
We needed to make sure that the steady state photon number in the lasing state is small enough that the resulting coherent state could be represented with only $P=10$ Fock states.
It is likely that, for regimes where the photon number is larger, the required value of $N$ to see convergence between these different approaches also increases.
This is supported by the results outlined in \ref{app:2QM} where the 2-level system shows similar agreement for an equivalent set of parameter choices.

\begin{figure}
\captionsetup{width=.9\linewidth}
  \lineskip=-\fboxrule
  \fbox{\begin{minipage}{\dimexpr \textwidth-2\fboxsep-2\fboxrule}
    \centering
    \includegraphics[height=11. cm, angle=0]{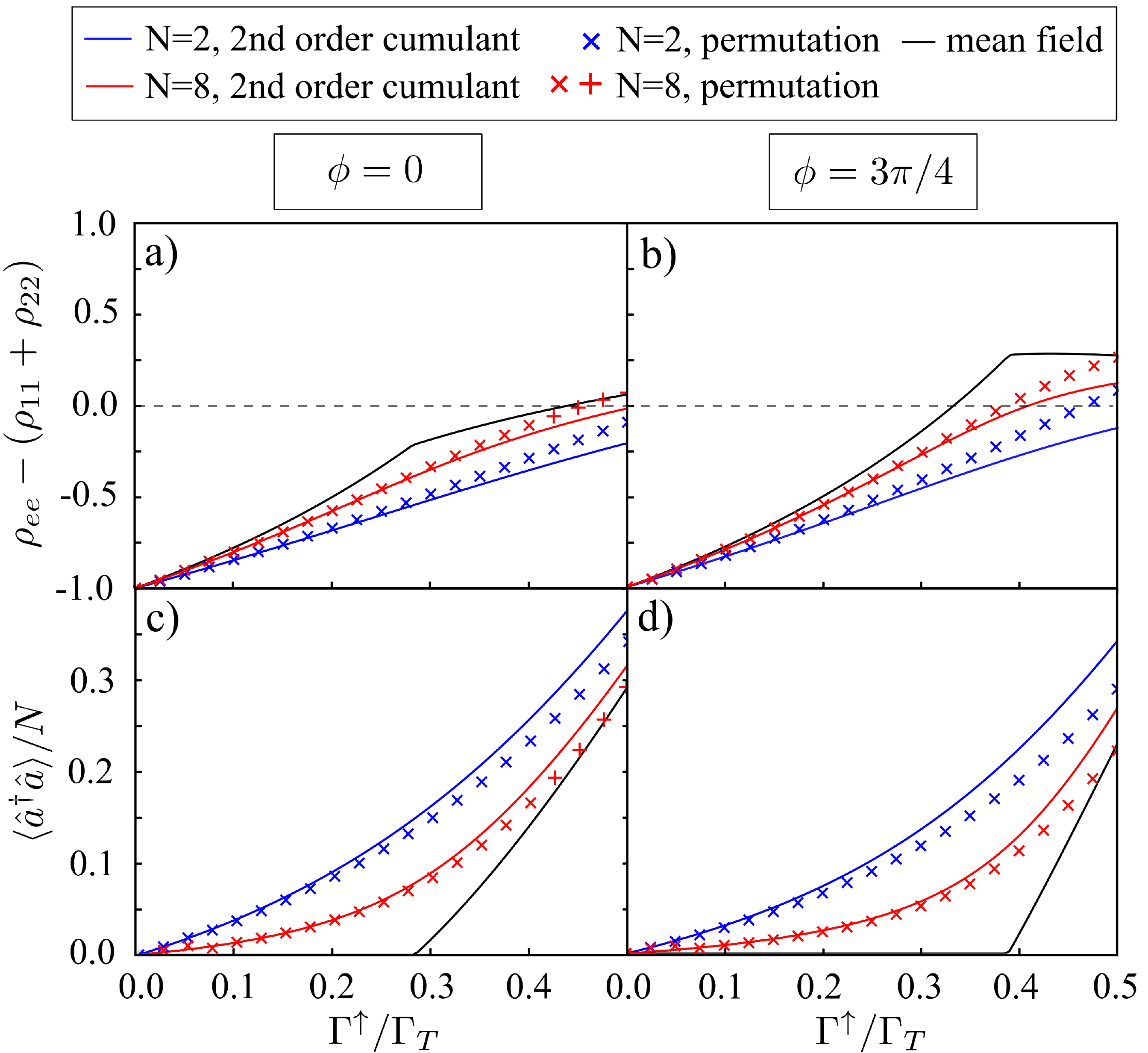}
    \abovecaptionskip=0pt
    \caption{Steady state solutions generated by the mean-field equations (solid black line) vs.~second order cumulant (solid coloured lines) vs.~exact numerical solutions calculated using permutation symmetry (crosses and plus-signs), the  pluses indicate where the truncated Fock space, limited to $P=10$, is no longer large enough to accurately describe the distribution of excitations in the cavity. All other parameters are the same as in Figure~\ref{fig:methodcomp1}.}
    \label{fig:InvCavPerm}
  \end{minipage}}
\end{figure}

Sweeping the pumping rate through the lasing transition, Figure~\ref{fig:InvCavPerm} provides a comparison between the steady state solutions for population inversion and cavity population calculated from the second order cumulant terms and permutation approach for $N=2$ and $N=8$.
As before, mean-field theory provides comparison for these in the $N\rightarrow\infty$ limit.
The two distinct cases $\phi = 0$ and $\phi = 3\pi/4$ where the transition is below and above inversion are explored in order to demonstrate that this effect can be seen using all of our approaches.
We again see that, even for small values of $N$ the agreement between the exact solutions and the cumulant expansion is very good, only breaking down slightly when the cavity occupation is large.
This is region where the atom-cavity interaction has the most significant effect meaning that higher order cumulants can be generated.
In every case shown by Figure~\ref{fig:InvCavPerm} we see the phase transition sharpening as $N$ increases with the discontinuity only apparent in the mean-field results.
We note that, Figures \ref{fig:InvCavPerm}a) and b) include data points in which the Fock space truncation $P=10$ is a limiting factor at the largest pumping strengths shown.
This slightly reduces the observed photon occupation compared to the exact value, as such we have indicated when the occupation of the highest energy Fock state exceeds $1/1000$.
This is most apparent at high pumping with $\phi=0$ which allows for a large cavity occupation.

\begin{figure}
\captionsetup{width=.9\linewidth}
  \lineskip=-\fboxrule
  \fbox{\begin{minipage}{\dimexpr \textwidth-2\fboxsep-2\fboxrule}
    \centering
    \includegraphics[height=11. cm, angle=0]{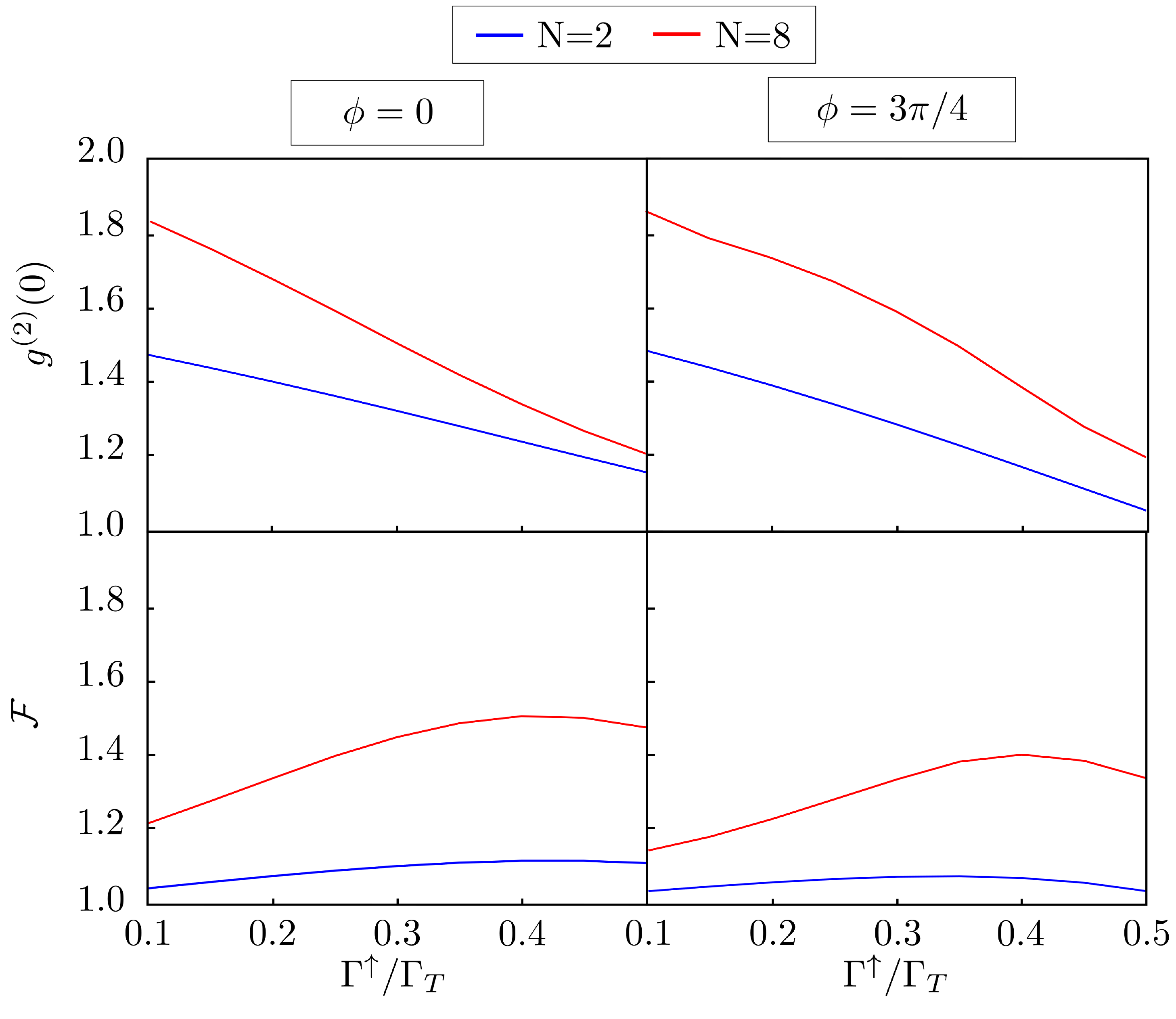}
    \abovecaptionskip=0pt
    \caption{Exact numerical simulations for the second order coherence $g^{(2)}(0)$ and Fano factor $\mathcal{F}$ as the system is pumped through the lasing phase transition for two different values of $\phi$. All parameters are the same as in Figure~\ref{fig:methodcomp1}.}
    \label{fig:PhotStatPerm}
  \end{minipage}}
\end{figure} 

To examine the nature of the light emitted by the cavity it is useful to examine higher order correlation functions of the photon distribution. These calculations are only straightforward for the exact results and so we are limited to the small numbers of emitters which are possible with the permutation approach.  In Fig.~\ref{fig:PhotStatPerm} we show both the second order coherence
\begin{equation}\label{eq:g2}
    g^{(2)}(0)= \frac{\langle a^\dagger a^\dagger aa\rangle}{\langle a^\dagger a\rangle^2},
\end{equation}
and Fano factor
\begin{equation}\label{eq:fano}
    \mathcal{F} = \frac{\langle a^\dagger a a^\dagger a \rangle - \langle a^\dagger a \rangle^2}{\langle a^\dagger a \rangle}.
\end{equation}
As the system is driven through the lasing transition we expect both of the quantities to show signatures of the change in the state of the radiation. The second order coherence reduces from $g^{(2)}(0)=2$ which is characteristic of thermal light to $g^{(2)}(0)=1$ which is characteristic of coherent light~\cite{ScullyBook}. The change between these two states becomes sharper as the thermodynamic limit is reached. The Fano factor in contrast is a measure of the scale of the fluctuations in the photon number, as such it shows a distinct peak at the transition point which corresponds to the large photon number uncertainty at this point. 
We observe the expected trend for both of these effects in the results for the 3-level systems, with the signatures becoming clearer for $N=8$. Our results for the 2-level model in~\ref{app:2QM}, where we are able to go to much larger $N$, are even more pronounced and clearly approach the expectations for the thermodynamic limit.

\section{Conclusion} \label{sec:conc}

We have constructed a Hamiltonian for a $\Lambda$-type 3-level system where the introduction of a microwave field provides a fully quantum model of the lasers originally described in Reference~\cite{Fearn92}.
We explicitly demonstrate how the interplay between the microwave dressed states and the light-matter interaction can lead to LWI for the correct choice of phase.

We characterised the system dynamics through mean-field equations, the solutions of which give the exact behaviour in the thermodynamic limit $N\to\infty$, and demonstrate how the location of the minimum in the lasing phase diagram is detuned by an amount $\Delta \omega\approx\Omega$ as a result of the introduction of the microwave field.
Linear stability analysis allowed us to derive a complete analytical expression for the location of the phase transition at certain values of the phase of the microwave field, $\phi$.
We show how the properties of the lasing transition can be optimised by adjusting this detuning along with the microwave  phase.

We go beyond mean-field theory to examine the behaviour of the system at finite system sizes. To do this we use two approaches: a cumulant expansion which gives access to corrections of order $1/N$ and exact numerical solutions, calculated using permutation symmetry which allow us to exactly find the steady state density operator of the system at small and intermediate numbers of 3-level systems. 
The combination of these approaches lets us characterise the properties of the lasing transition across the whole range of possible system sizes with good agreement in the expected regimes.
For the parameters chosen here we were able to see agreement between the cumulant expansion and exact numerics even at $N=8$.

The microwave field described here introduces a degree of quantum control over the laser phase transition of the system.
Furthermore, the framework outlined would provide a genuine advantage for understanding intermediate $N$ systems where cross-correlations accurately captured by the cumulant expansion have a significant impact upon steady state behaviour.
Such systems, with a bespoke architecture of energy levels, are at the heart of quantum technologies beyond lasing such as atomic clocks.
Therefore, with the quantum control demonstrated here alongside a rich analysis of finite sized systems, there is great potential to build upon this work by exploring further quantum systems which can be experimentally realised through nanophotonics.

\section*{Acknowledgements}

NW and EMG acknowledge funding from the EPSRC grant no.~EP/T007214/1.

\section*{References}
\bibliographystyle{unsrt}
\bibliography{main}

\begin{thebibliography}{10}

\bibitem{Haken70}
H.~Haken.
\newblock In S.~M. Kay and A.~Maitland, editors, {\em Quantum Optics}, page
  201. Academic Press, London, 1970.

\bibitem{ScullyBook}
M.~O. Scully and M.~S. Zubairy.
\newblock {\em Quantum Optics}.
\newblock Cambridge University Press, 1997.

\bibitem{Scully89}
M.~O. Scully, S.~Y. Zhu, and A.~Gavrielides.
\newblock Degenerate quantum-beat laser: Lasing without inversion and inversion
  without lasing.
\newblock {\em Phys. Rev. Lett.}, 62:2813, 1989.

\bibitem{Scully92}
M.~O. Scully, S.~Y. Zhu, and H.~Fearn.
\newblock Lasing without inversion {I}: Initial atomic coherence.
\newblock {\em Z. Phys. D}, 22:471, 1992.

\bibitem{Zhu92}
S.~Y. Zhu, M.~O. Scully, H.~Fearn, and L.~M. Narducci.
\newblock Lasing without inversion {II}: Raman process created atomic
  coherence.
\newblock {\em Z. Phys. D}, 22:483, 1992.

\bibitem{Gray78}
H.~R. Gray, R.~M. Whitley, and C.~R. Stroud.
\newblock Coherent trapping of atomic populations.
\newblock {\em Opt. Lett.}, 3:218, 1978.

\bibitem{Xu08}
X.~Xu, B.~Sun, P.~R. Berman, D.~G. Steel, A.~S. Bracker, D.~Gammon, and L.~J.
  Sham.
\newblock Coherent population trapping of an electron spin in a single
  negatively charged quantum dot.
\newblock {\em Nat. Phys.}, 4:692, 2008.

\bibitem{Gaubatz90}
U.~Gaubatz, P.~Rudecki, S.~Schiemann, and K.~Bergmann.
\newblock Population transfer between molecular vibrational levels by
  stimulated raman scattering with partially overlapping laser fields. a new
  concept and experimental results.
\newblock {\em J. Chem. Phys.}, 92:5363, 1990.

\bibitem{Kumar16}
K.~S. Kumar, A.~Veps\"{a}l\"{a}inen, S.~Danilin, and G.~S. Paraoanu.
\newblock Stimulated raman adiabatic passage in a three-level superconducting
  circuit.
\newblock {\em Nat. Commun.}, 7:10628, 2016.

\bibitem{Boller91}
K.~J. Boller, A.~Imamo\ifmmode~\breve{g}\else \u{g}\fi{}lu, and S.~E. Harris.
\newblock Observation of electromagnetically induced transparency.
\newblock {\em Phys. Rev. Lett.}, 66:2593, 1991.

\bibitem{Naeini11}
A.~H. Safavi-Naeini, T.~P. Mayer~Alegre, J.~Chan, M.~Eichenfield, M.~Winger,
  Q.~Lin, J.~T. Hill, D.~E. Chang, and O.~Painter.
\newblock Electromagnetically induced transparency and slow light with
  optomechanics.
\newblock {\em Nature}, 472:69, 2011.

\bibitem{Vanier05}
J.~Vanier.
\newblock Atomic clocks based on coherent population trapping: a review.
\newblock {\em Applied Physics B}, 81:421, 2005.

\bibitem{Santra05}
R.~Santra, E.~Arimondo, T.~Ido, C.~H. Greene, and J.~Ye.
\newblock High-accuracy optical clock via three-level coherence in neutral
  bosonic $^{88}\mathrm{Sr}$.
\newblock {\em Phys. Rev. Lett.}, 94:173002, 2005.

\bibitem{Zanon14}
T.~Zanon-Willette, S.~Almonacil, E.~Clercq, A.~Ludlow, and E.~Arimondo.
\newblock Quantum engineering of atomic phase-shifts in optical clocks.
\newblock {\em Phys. Rev. A}, 90:053427, 2014.

\bibitem{Pedrozo20}
E.~Pedrozo-Pe{\~n}afiel, S.~Colombo, C.~Shu, A.~F. Adiyatullin, Z.~Li,
  E.~Mendez, B.~Braverman, A.~Kawasaki, D.~Akamatsu, Y.~Xiao, and V.~Vuleti\'c.
\newblock Entanglement on an optical atomic-clock transition.
\newblock {\em Nature}, 588:414, 2020.

\bibitem{Fleischhauer02}
M.~Fleischhauer and M.~D. Lukin.
\newblock Quantum memory for photons: Dark-state polaritons.
\newblock {\em Phys. Rev. A}, 65:022314, 2002.

\bibitem{Lvovsky09}
A.~I. Lvovsky, B.~C. Sanders, and W.~Tittel.
\newblock Optical quantum memory.
\newblock {\em Nat. Photonics}, 3:706, 2009.

\bibitem{Aspect88}
A.~Aspect, E.~Arimondo, R.~Kaiser, N.~Vansteenkiste, and C.~Cohen-Tannoudji.
\newblock Laser cooling below the one-photon recoil energy by
  velocity-selective coherent population trapping.
\newblock {\em Phys. Rev. Lett.}, 61:826, 1988.

\bibitem{Marzoli94}
I.~Marzoli, J.~I. Cirac, R.~Blatt, and P.~Zoller.
\newblock Laser cooling of trapped three-level ions: Designing two-level
  systems for sideband cooling.
\newblock {\em Phys Rev A}, 49:2771, 1994.

\bibitem{Morigi00}
G.~Morigi, J.~Eschner, and C.~H. Keitel.
\newblock Ground state laser cooling using electromagnetically induced
  transparency.
\newblock {\em Phys. Rev. Lett.}, 85:4458, 2000.

\bibitem{Zhou02}
S.~Zhou, S.~I. Chu, and S.~Han.
\newblock Quantum computing with superconducting devices: A three-level squid
  qubit.
\newblock {\em Phys. Rev. B}, 66:054527, 2002.

\bibitem{Rao14}
D.~D.~B. Rao and K.~M{\o}lmer.
\newblock Robust {R}ydberg-interaction gates with adiabatic passage.
\newblock {\em Phys. Rev. A}, 89:030301, 2014.

\bibitem{Higgins17}
G.~Higgins, F.~Pokorny, C.~Zhang, Q.~Bodart, and M.~Hennrich.
\newblock Coherent control of a single trapped {R}ydberg ion.
\newblock {\em Phys. Rev. Lett.}, 119:220501, 2017.

\bibitem{Azzam20}
S.~I. Azzam, A.~V. Kildishev, R.~M. Ma, C.~Z. Ning, R.~Oulton, V.~M. Shalaev,
  M.~I. Stockman, J.~L. Xu, and X.~Zhang.
\newblock Ten years of spasers and plasmonic nanolasers.
\newblock {\em Light Sci. Appl.}, 9:90, 2020.

\bibitem{Breeze18}
J.~D. Breeze, E.~Salvadori, J.~Sathian, N.~McN. Alford, and C.~W.~M. Kay.
\newblock Continuous-wave room-temperature diamond maser.
\newblock {\em Nature}, 555:493, 2018.

\bibitem{Ning19}
C.~Z. Ning.
\newblock {Semiconductor nanolasers and the size-energy-efficiency challenge: a
  review}.
\newblock {\em Adv. Photonics}, 1:014002, 2019.

\bibitem{Protsenko05}
I.~E. Protsenko, A.~V. Uskov, O.~A. Zaimidoroga, V.~N. Samoilov, and E.~P.
  O'Reilly.
\newblock Dipole nanolaser.
\newblock {\em Phys. Rev. A}, 71:063812, 2005.

\bibitem{Rice94}
P.~R. Rice and H.~J. Carmichael.
\newblock Photon statistics of a cavity-qed laser: A comment on the
  laser--phase-transition analogy.
\newblock {\em Phys. Rev. A}, 50:4318, 1994.

\bibitem{Nomura09}
M.~Nomura, N.~Kumagai, S.~Iwamoto, Y.~Ota, and Y.~Arakawa.
\newblock Photonic crystal nanocavity laser with a single quantum dot gain.
\newblock {\em Opt. Express}, 17:15975, 2009.

\bibitem{Ojambati21}
O.~S. Ojambati, K.~B. Arnardottir, B.~W. Lovett, J.~Keeling, and J.~J.
  Baumberg.
\newblock Few-emitter lasing in single ultra-small nanocavities.
\newblock {\em Arxiv preprint 2107.14304}, 2021.

\bibitem{Slussarenko19}
S.~Slussarenko and G.~J. Pryde.
\newblock Photonic quantum information processing: A concise review.
\newblock {\em App. Phys. Rev.}, 6:041303, 2019.

\bibitem{Leymann15}
H.~A.~M. Leymann, A.~Foerster, F.~Jahnke, J.~Wiersig, and C.~Gies.
\newblock Sub- and superradiance in nanolasers.
\newblock {\em Phys. Rev. Applied}, 4:044018, 2015.

\bibitem{Andre2019}
E.~C. Andr{\'e}, I.~E. Protsenko, A.~V. Uskov, J.~M{\o}rk, and M.~Wubs.
\newblock On collective {R}abi splitting in nanolasers and nano-{LED}s.
\newblock {\em Optics letters}, 44:1415, 2019.

\bibitem{Mork20}
J.~Mork and K.~Yvind.
\newblock Squeezing of intensity noise in nanolasers and nano{LED}s with
  extreme dielectric confinement.
\newblock {\em Optica}, 7:1641, 2020.

\bibitem{Javan57}
A.~Javan.
\newblock Theory of a three-level maser.
\newblock {\em Phys. Rev.}, 107:1579, 1957.

\bibitem{Mompart00}
J.~Mompart and R.~Corbal{\'{a}}n.
\newblock Lasing without inversion.
\newblock {\em J. Opt. B: Quant. Semiclass. Opt.}, 2:R7, 2000.

\bibitem{Richter20}
M.~Richter, M.~Lytova, F.~Morales, S.~Haessler, O.~Smirnova, M.~Spanner, and
  M.~Ivanov.
\newblock Rotational quantum beat lasing without inversion.
\newblock {\em Optica}, 7:586, 2020.

\bibitem{Marcuse63}
D.~Marcuse.
\newblock Maser action without population inversion.
\newblock {\em (BTL) Proc. IEEE}, 51:849, 1963.

\bibitem{Holt77}
H.~K. Holt.
\newblock Gain without population inversion in two-level atoms.
\newblock {\em Phys. Rev. A}, 16:1136, 1977.

\bibitem{Fearn92}
H.~Fearn, M.~O. Scully, S.~Y. Zhu, and M.~Sargent.
\newblock Lasing without inversion {III}: microwave coupling induced atomic
  coherence.
\newblock {\em Z. Phys. D}, 22:495, 1992.

\bibitem{Kocharovskaya92}
O.~Kocharovskaya, P.~Mandel, and Y.~V. Radeonychev.
\newblock Inversionless amplification in a three-level medium.
\newblock {\em Phys. Rev. A}, 45:1997, 1992.

\bibitem{Scully87}
M.~O. Scully and M.~S. Zubairy.
\newblock Theory of the quantum-beat laser.
\newblock {\em Phys. Rev. A}, 35:752, 1987.

\bibitem{Zibrov95}
A.~S. Zibrov, M.~D. Lukin, D.~E. Nikonov, L.~Hollberg, M.~O. Scully, V.~L.
  Velichansky, and H.~G. Robinson.
\newblock Experimental demonstration of laser oscillation without population
  inversion via quantum interference in {Rb}.
\newblock {\em Phys. Rev. Lett.}, 75:1499, 1995.

\bibitem{Peters96}
C.~Peters and W.~Lange.
\newblock Laser action below threshold inversion due to coherent population
  trapping.
\newblock {\em Appl. Phys. B}, 62:221, 1996.

\bibitem{Lin21}
R.~Lin, R.~Rosa-Medina, F.~Ferri, F.~Finger, K.~Kroeger, T.~Donner,
  T.~Esslinger, and R.~Chitra.
\newblock Dissipation-engineered family of nearly dark states in many-body
  cavity-atom systems.
\newblock {\em Phys. Rev. Lett.}, 128:153601, 2022.

\bibitem{Minganti21}
F.~Minganti, I.~I. Arkhipov, A.~Miranowicz, and F.~Nori.
\newblock Liouvillian spectral collapse in scully-lamb lasing: Non-equilibrium
  second-order phase transition with or without {$U(1)$} symmetry breaking.
\newblock {\em Arxiv preprint 2103.05625}, 2021.

\bibitem{Doronin19}
I.~V. Doronin, A.~A. Zyablovsky, E.~S. Andrianov, A.~A. Pukhov, and A.~P.
  Vinogradov.
\newblock Lasing without inversion due to parametric instability of the laser
  near the exceptional point.
\newblock {\em Phys. Rev. A}, 100:021801, 2019.

\bibitem{Miri19}
M.~A. Miri and A.~Al\'{u}.
\newblock Exceptional points in optics and photonics.
\newblock {\em Science}, 363:6422, 2019.

\bibitem{Richter15}
M.~Richter, M.~Gegg, T.~S. Theuerholz, and A.~Knorr.
\newblock Numerically exact solution of the many emitter--cavity laser problem:
  Application to the fully quantized spaser emission.
\newblock {\em Phys. Rev. B}, 91:035306, 2015.

\bibitem{Kirton17}
P.~Kirton and J.~Keeling.
\newblock Suppressing and restoring the {D}icke superradiance transition by
  dephasing and decay.
\newblock {\em Phys. Rev. Lett.}, 118:123602, 2017.

\bibitem{Shammah18}
N.~Shammah, S.~Ahmed, N.~Lambert, S.~De~Liberato, and F.~Nori.
\newblock Open quantum systems with local and collective incoherent processes:
  Efficient numerical simulations using permutational invariance.
\newblock {\em Phys. Rev. A}, 98:063815, 2018.

\bibitem{Kirton19}
P.~Kirton, M.~M. Roses, J.~Keeling, and E.~G. Dalla~Torre.
\newblock Introduction to the {D}icke model: From equilibrium to
  nonequilibrium, and vice versa.
\newblock {\em Adv. Quantum Technol.}, 2:1800043, 2018.

\bibitem{Wang73}
Y.~K. Wang and F.~T. Hioe.
\newblock Phase transition in the {D}icke model of superradiance.
\newblock {\em Phys. Rev. A}, 7:831, 1973.

\bibitem{Emary03}
C.~Emary and T.~Brandes.
\newblock Chaos and the quantum phase transition in the {D}icke model.
\newblock {\em Phys. Rev. E}, 67:066203, 2003.

\bibitem{Wiseman09}
H.~M. Wiseman and G.~J. Milburn.
\newblock {\em Quantum Measurement and Control}.
\newblock Cambridge University Press, 2009.

\bibitem{Gothe2019}
H.~Gothe, D.~Sholokhov, A.~Breunig, M.~Steinel, and J.~Eschner.
\newblock Continuous-wave virtual-state lasing from cold ytterbium atoms.
\newblock {\em Phys. Rev. A}, 99:013415, 2019.

\bibitem{Thomas2019}
S.~E. Thomas, T.~M. Hird, J.~H.~D. Munns, B.~Brecht, D.~J. Saunders, J.~Nunn,
  I.~A. Walmsley, and P.~M. Ledingham.
\newblock Raman quantum memory with built-in suppression of four-wave-mixing
  noise.
\newblock {\em Phys. Rev. A}, 100:033801, Sep 2019.

\bibitem{Kutluer2019}
K.~Kutluer, E.~Distante, B.~Casabone, S.~Duranti, M.~Mazzera, and
  H.~de~Riedmatten.
\newblock Time entanglement between a photon and a spin wave in a multimode
  solid-state quantum memory.
\newblock {\em Phys. Rev. Lett.}, 123:030501, 2019.

\bibitem{Kubo62}
R.~Kubo.
\newblock Generalized cumulant expansion method.
\newblock {\em J. Phys. Soc. Japan}, 17:1100, 1962.

\bibitem{Plankensteiner22}
D.~Plankensteiner, C.~Hotter, and H.~Ritsch.
\newblock Quantum{C}umulants.jl: {A} {J}ulia framework for generalized
  mean-field equations in open quantum systems.
\newblock {\em {Quantum}}, 6:617, 2022.

\bibitem{KirtonGit17}
P.~Kirton.
\newblock Peterkirton/permutations:permutationsv1.0
  (https://doi.org/10.5281/zenodo.376621), 2017.

\bibitem{Chase08}
B.~A. Chase and J.~M. Geremia.
\newblock Collective processes of an ensemble of spin-$1/2$ particles.
\newblock {\em Phys. Rev. A}, 78:052101, 2008.

\bibitem{Damanet16}
F.~Damanet, D.~Braun, and J.~Martin.
\newblock Cooperative spontaneous emission from indistinguishable atoms in
  arbitrary motional quantum states.
\newblock {\em Phys. Rev. A}, 94:033838, 2016.

\bibitem{Gegg18}
M.~Gegg, A.~Carmele, A.~Knorr, and M.~Richter.
\newblock Superradiant to subradiant phase transition in the open system
  {D}icke model: dark state cascades.
\newblock {\em New J. Phys.}, 20:013006, 2018.

\bibitem{Kirton18}
P.~Kirton and J.~Keeling.
\newblock Superradiant and lasing states in driven-dissipative {D}icke models.
\newblock {\em New J. Phys.}, 20:015009, 2018.

\bibitem{Wang21}
P.~Wang and R.~Fazio.
\newblock Dissipative phase transitions in the fully connected {I}sing model
  with p-spin interaction.
\newblock {\em Phys. Rev. A}, 103:013306, 2021.

\bibitem{Gartner2011}
P.~Gartner.
\newblock Two-level laser: Analytical results and the laser transition.
\newblock {\em Phys. Rev. A}, 84:053804, 2011.

\end{thebibliography}

\appendix
\section{2-Level System: Model} \label{app:2QM}

{For comparison, and to highlight the features that uniquely arise from the microwave-driven 3-level $\Lambda$ structure, we here provide the analogous results for a collection of incoherently pumped 2-level systems coupled to a lossy cavity mode.}
Our approach follows previous work, specifically the textbook Dicke and Scully-Lamb laser model~\cite{ScullyBook, Haken70}.
{Here we use -- to the extent possible -- the same parameters and notation as in the main text to aid with this comparison}.
We again describe the cavity using a single mode approximation
\begin{equation} \label{eq:HF}
    H_{F} = \nu  a^{\dag} a.
\end{equation}
The collection of 2-level atomic systems and their interaction with the cavity are described by the following Hamiltonian
\begin{equation} \label{eq:HA}
H^{(i)}_{AF} = \omega_{e} \rho^{(i)}_{ee} + \frac{g}{\sqrt{N}}\left( a^{\dag} \rho^{(i)}_{ge} + a \rho^{(i)}_{eg}  \right),
\end{equation}
where we again use the energy of the ground state as our reference $\omega_g=0$ and $\rho_{jk}^{(i)} = \ket{j} \bra{k}^{(i)}$ is an operator on the $i$th 2-level system.
In the atom-field Hamiltonian, we make the rotating wave approximation, discarding the counter-rotating terms, and assume $g$ is real. 
As such, the resulting quantum master equation for $N$ systems is {[\textit{c.f.}~Equation (\ref{eq:QME1})]}:
\begin{equation}
    \dot{\rho}(t) = -i\left[ H, \rho(t)\right] + \kappa \mathcal{D}[a] +\sum_i \Gamma^{\uparrow} \mathcal{D}[\rho^{(i)}_{eg}] + \Gamma^{\downarrow} \mathcal{D}[\rho^{(i)}_{ge}] ,
\end{equation}
where 
\begin{equation}
    H = H_{F} + \sum_i H^{(i)}_{AF}.
\end{equation}
This model again has a $U(1)$ symmetry corresponding to the phase of the cavity field operator and atomic coherence which is spontaneously broken at the lasing phase transition.

\begin{figure}
\captionsetup{width=.5\linewidth}
\centering
  \lineskip=-\fboxrule
  \fbox{\begin{minipage}[t]{0.7\linewidth}
    \centering
    \includegraphics[height=5.0 cm, angle=0]{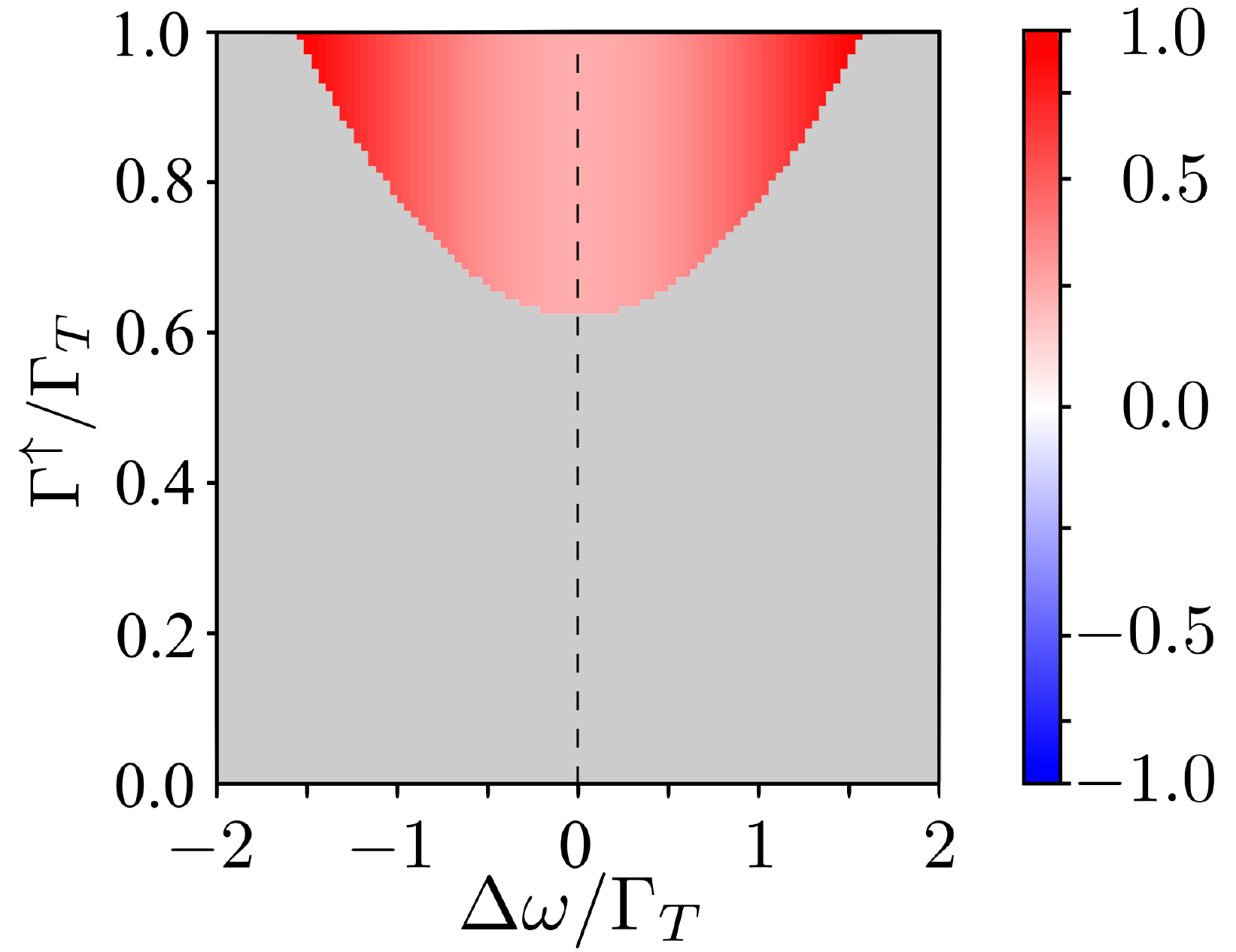}
    \abovecaptionskip=0pt
    \caption{Laser phase transition for an ensemble of 2-level systems interacting with a single mode cavity. The magnitude of inversion associated with the lasing region is indicated by the shading. Excluding those labelled, the parameters are $\omega_{e}=\Gamma_{T}$, $g=0.9\Gamma_{T}$, and $\kappa = 0.8\Gamma_{T}$.}
    \label{fig:2LSLaser}
  \end{minipage}}
\end{figure} 

\begin{figure}
\captionsetup{width=.9\linewidth}
  \lineskip=-\fboxrule
  \fbox{\begin{minipage}{\dimexpr \textwidth-2\fboxsep-2\fboxrule}
    \centering
    \includegraphics[height=9 cm, angle=0]{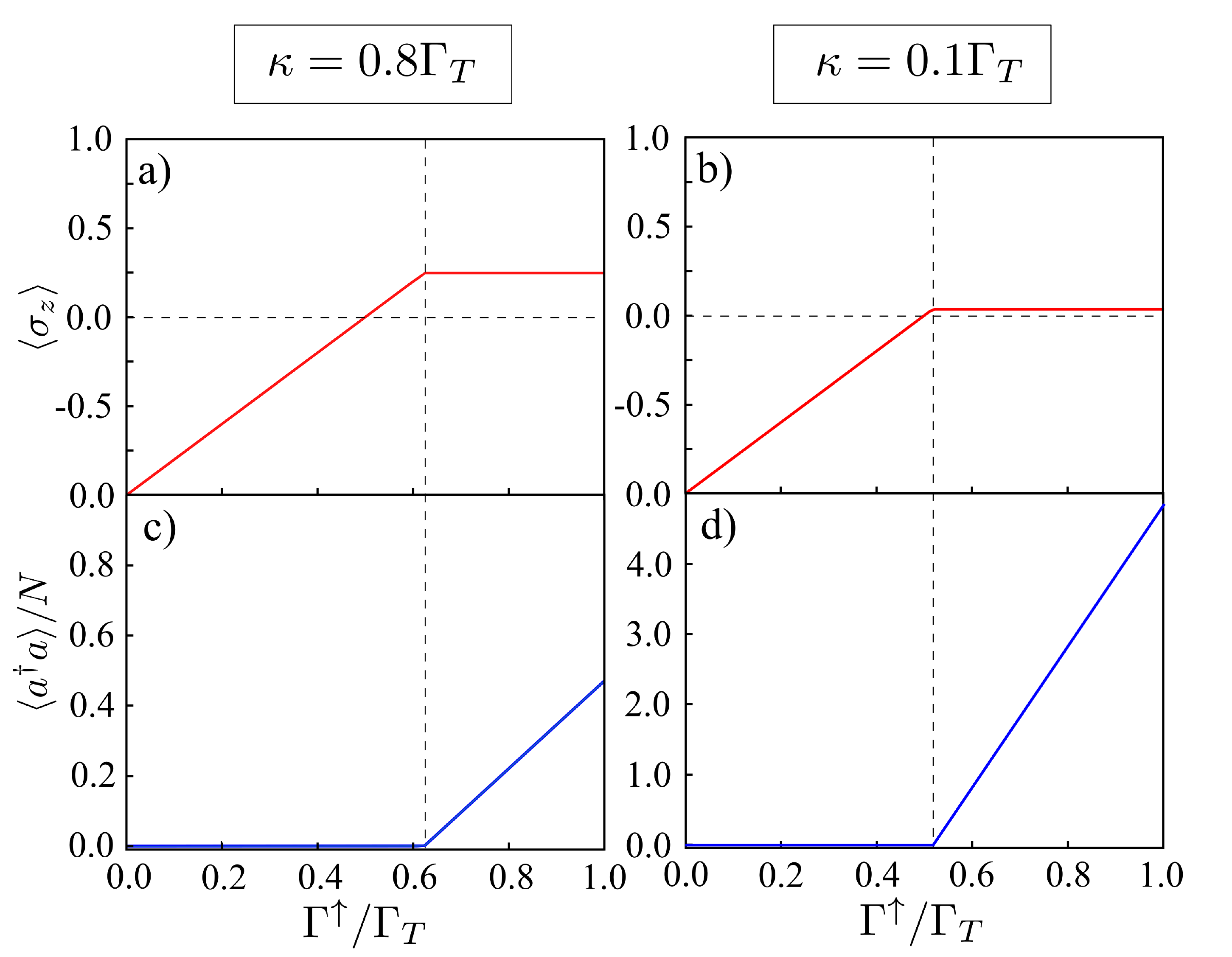}
    \abovecaptionskip=0pt
    \caption{Atomic population inversion and cavity population as a function of pumping strength for different values of cavity decay rate. Upper row: Population inversion at the indicated values of $\Gamma^{\uparrow}/\Gamma_{T}$ and $\kappa$. The horizontal dashed line shows the point at which the excited and ground state populations are equal, whereas the vertical dashed lines indicate the analytically calculated point at which the lasing phase transition occurs. Lower row: Cavity population for the same parameters. Excluding those labelled, the parameters are the same as Figure~\ref{fig:2LSLaser} and $\Delta \omega = 0$. }
    \label{fig:2LSAtomPop}
  \end{minipage}}
\end{figure}

As with the 3-level system, we construct a set of mean-field equations for the operators of the 2-level system which are as follows:
\begin{gather}
    \braket{\dot{\rho}_{ge}} =  -i\left(\omega_{e}\braket{\rho_{ge}} -\frac{g}{\sqrt{N}} \braket{a} \braket{\sigma_{z}}\right) - \frac{\Gamma_{T}}{2} \braket{\rho_{ge}}, \\
    \braket{\dot\sigma_{z}} =-\frac{2g}{\sqrt{N}}\left( i\braket{a} \braket{\rho_{eg}} + \text{c.c.} \right) + \Gamma_{T} \braket{\sigma_{z}} + (\Gamma^{\uparrow} - \Gamma^{\downarrow}),
    \\
    \braket{\dot{a}} = -i\left( \nu\braket{a} +g\sqrt{N}\rho_{ge}\right) - \frac{\kappa}{2}\braket{a},
\end{gather}
where $\Gamma_{T} = \Gamma^{\uparrow} + \Gamma^{\downarrow}$, $\sigma_{z} = \rho_{ee} - \rho_{gg}$, and $\braket{\rho_{ee}} + \braket{\rho_{gg}} = 1$.
From the mean-field equations, linear stability analysis of the normal state can be used to further understand the phase transition.
The normal state is always defined such that $\braket{a}_{ns}=\braket{\rho_{eg}}_{ns} = 0$ and $\braket{\sigma_z}_{ns}=(\Gamma^{\uparrow} - \Gamma^{\downarrow})/\Gamma_{T}$. 
Linearising around this solution, such that $\delta a = \braket{a} - \braket{a}_{ns}$, $\delta \rho_{eg} = \braket{\rho_{eg}} - \braket{\rho_{eg}}_{ns}$ and $\delta \sigma_z = \braket{\sigma_z} - \braket{\sigma_z}_{ns}$ the resulting matrix describes the linear stability of the normal state is,
\begin{equation}
\frac{d}{dt}
\begin{pmatrix}
\delta a\\
\delta \rho_{ge}\\
\delta \sigma_z 
\end{pmatrix}
=
\begin{pmatrix}
-i \nu - \frac{\kappa}{2} & -ig\sqrt{N} & 0\\
i \frac{g}{\sqrt{N}} \braket{\sigma_{z}}_{ns} & -i\omega_{e} - \frac{\Gamma_T}{2}  & 0\\
0 & 0 & -\Gamma_{T}
\end{pmatrix}\begin{pmatrix}
\delta a\\
\delta \rho_{ge}
\\
\delta \sigma_z 
\end{pmatrix}.
\end{equation}
We note that to find the phase transition from this matrix we do not need to consider the fluctuations of $\delta\sigma_z$ since this decouples from the other quantities and is always stable.
By calculating the characteristic polynomial of this matrix, 
\begin{equation}
   \lambda^{2} + \left[\frac{\Gamma_{T} + \kappa}{2} - i(\omega_{e} + \nu) \right] \lambda + g^{2}\frac{\left(\Gamma^{\uparrow} - \Gamma^{\downarrow} \right)}{\Gamma_{T}} + \left( \frac{\Gamma_{T}}{2} - i\omega_{e} \right)\left(\frac{\kappa}{2} - i\nu \right)=0,
\end{equation}
the critical pumping rate at the lasing phase transition is found to be
\begin{equation}
   \Gamma^{\uparrow}_{\text{crit}} = \Gamma_{T}\left(\frac{1}{2} + \frac{\Gamma_{T} \kappa}{g^{2}} \left[\frac{1 }{8}+ \frac{(\Delta\omega)^{2}}{ (\Gamma_{T}+\kappa)^{2}} \right]\right),
\end{equation}
where $\Delta \omega = \omega_{e} - \nu$ is the atom-cavity detuning. 
It is clear that the resonance condition here is that the cavity should be at the same frequency as the atomic transition $\nu=\omega_e$. At this point and in the coherent limit of  $g\gg\kappa$ this reduces to $\Gamma^\uparrow_{\text{crit}}=\Gamma_T/2=\Gamma^\downarrow$. As expected the optimum lasing transition occurs exactly when gain and loss are balanced. This can be clearly seen in Figures~\ref{fig:2LSLaser} and \ref{fig:2LSAtomPop} where we show the phase diagram as a function of detuning and the steady state properties as we go through the lasing threshold. Notice that the gain locking shown in Figure~\ref{fig:2LSAtomPop}a)-b) is perfect~\cite{Rice94, Gartner2011}, in contrast to that observed in the 3-level system model.  We see that as the cavity decay rate is lowered the results approach the limit described above.

\subsection{Cumulant Equations}

\begin{figure}
\captionsetup{width=.9\linewidth}
  \lineskip=-\fboxrule
  \fbox{\begin{minipage}{\dimexpr \textwidth-2\fboxsep-2\fboxrule}
    \centering
    \includegraphics[height=9.0 cm, angle=0]{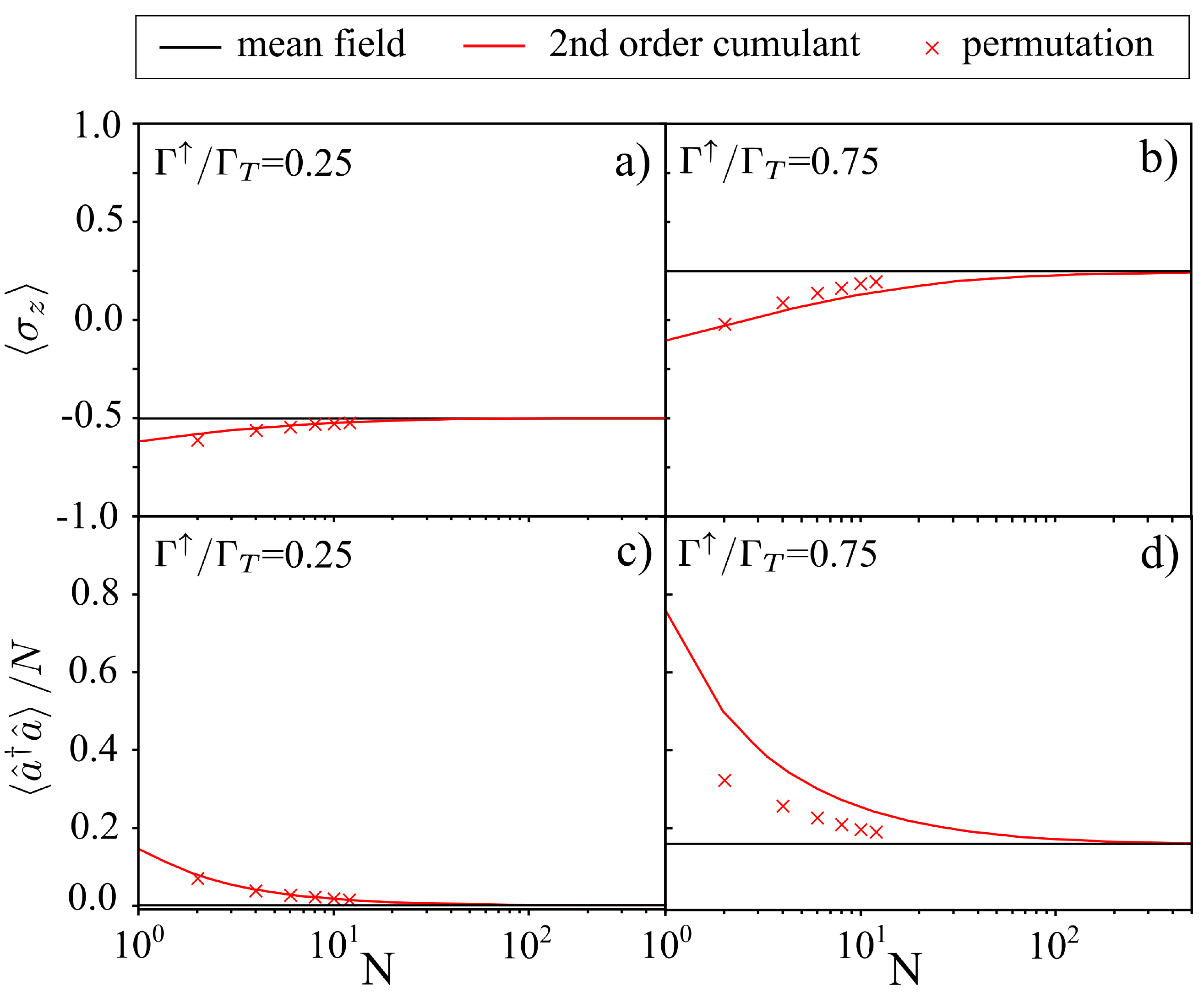}
    \abovecaptionskip=0pt
    \caption{The inversion and cavity population of the 2-level system calculated for a single pumping value either side of the phase transition. We compare how the second order cumulant (dashed line) and quantum model (crosses) converge on the mean-field equation (solid black line). Excluding those labelled, the parameters are the same as Figure \ref{fig:2LSAtomPop}.}
    \label{fig:ComparisonFig2} 
  \end{minipage}}
\end{figure}

\begin{figure}
\captionsetup{width=.9\linewidth}
  \lineskip=-\fboxrule
  \fbox{\begin{minipage}{\dimexpr \textwidth-2\fboxsep-2\fboxrule}
    \centering
    \includegraphics[height=6.0 cm, angle=0]{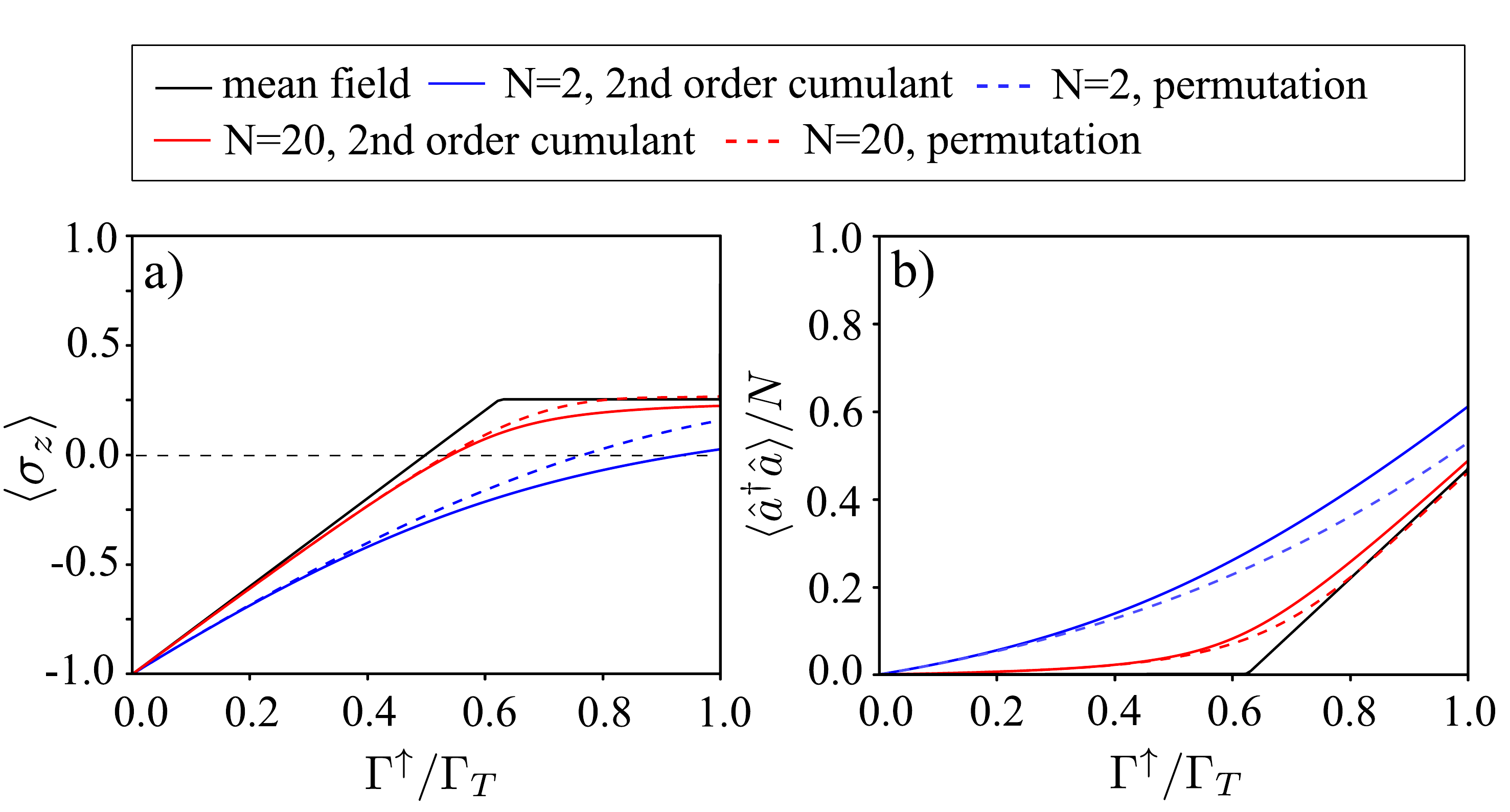}
    \abovecaptionskip=0pt
    \caption{Population inversion and cavity population of the 2-level system calculated from the mean-field equation (solid black line), second order cumulant (dashed lines) and quantum model (dotted lines). Excluding those labelled, the parameters are the same as Figure \ref{fig:2LSAtomPop}.}
    \label{fig:ComparisonFig1}
  \end{minipage}}
\end{figure}

The second order cumulants for 2-level system laser model provide a useful benchmark for the results of this paper.
The relevant equations for the matter terms obey the following
\begin{gather}
    \braket{\dot{\sigma_{z}}} =-\frac{2g}{\sqrt{N}}\left(i\braket{a\rho_{eg}} + \text{c.c.} \right) + \Gamma_{T} \braket{\sigma_{z}} + (\Gamma^{\uparrow} - \Gamma^{\downarrow}), \\
  \braket{\dot{\rho_{eg}\rho_{ge}}} =  - \Gamma_{T}\braket{\rho_{eg}\rho_{ge}} + i \frac{g}{\sqrt{N}} \left(\braket{a \rho_{eg}}\braket{\sigma_{z}} + \text{c.c.} \right).
\end{gather}
We have broken the third order moments that arise into products of first and second order moments by setting the third cumulants to zero.
 We choose to write the symmetry preserving versions of these equations such that we set all terms which break the $U(1)$ symmetry of the model to 0. This approach has been shown to give good agreement with exact results since the full quantum model always respects this symmetry for finite $N$~\cite{Plankensteiner22}.
Here the cross correlation terms denote operators acting on different atoms i.e.~$\braket{{\rho_{eg}\rho_{eg}}}\equiv \braket{{\rho_{eg}^{(i)}\rho_{eg}^{(j)}}}$ with $i\neq j$. Since, within the model we consider, all atoms are identical and hence all correlations between different pairs of atoms are identical we are able to drop the explicit labelling of these.
The second moments of the field-matter terms obey the following equations of motion:
\begin{multline}
   \braket{\dot{a\rho_{eg}}} = -\left[i\left(\nu - \omega_{e}\right) + \frac{1}{2}\left(\kappa + \Gamma_T\right)\right]\braket{a\rho_{eg}} -i\frac{g}{\sqrt{N}}\left[(N-1)\braket{\rho_{eg}\rho_{ge}} + \braket{a^{\dag}a}\braket{\sigma_{z}} \right. \\
\left. +\braket{\rho_{ee}}\right].
\end{multline}
Finally, the second moments of the field-field terms obey the following equations of motion:
\begin{gather}
    \braket{\dot{a^{\dag}a}} = - \kappa \braket{a^{\dag}a}  + g\sqrt{N}(i\braket{a\rho_{eg}} + \text{c.c.}).
\end{gather}
Together these provide a closed set of equations for the second moments which can be used to find the $1/N$ corrections to the mean-field equations.
A comparison between the behaviour described by the second order cumulants, mean-field equations, and the exact solution to the full quantum model obtained from the permutation symmetric method are shown in Figures~\ref{fig:ComparisonFig2} and \ref{fig:ComparisonFig1}. These again show how the exact solution approaches the cumulant results at relatively small $N$. The trend of moving towards the mean-field result at large $N$ can clearly be seen. These results should be compared to Figures~\ref{fig:methodcomp1} and \ref{fig:InvCavPerm} of the main text. We see very similar behaviour in both cases.

\begin{figure}
\captionsetup{width=0.9\linewidth}
\centering
  \lineskip=-\fboxrule
  \fbox{\begin{minipage}[t]{1.0\linewidth}
    \centering
    \includegraphics[height=6.0 cm, angle=0]{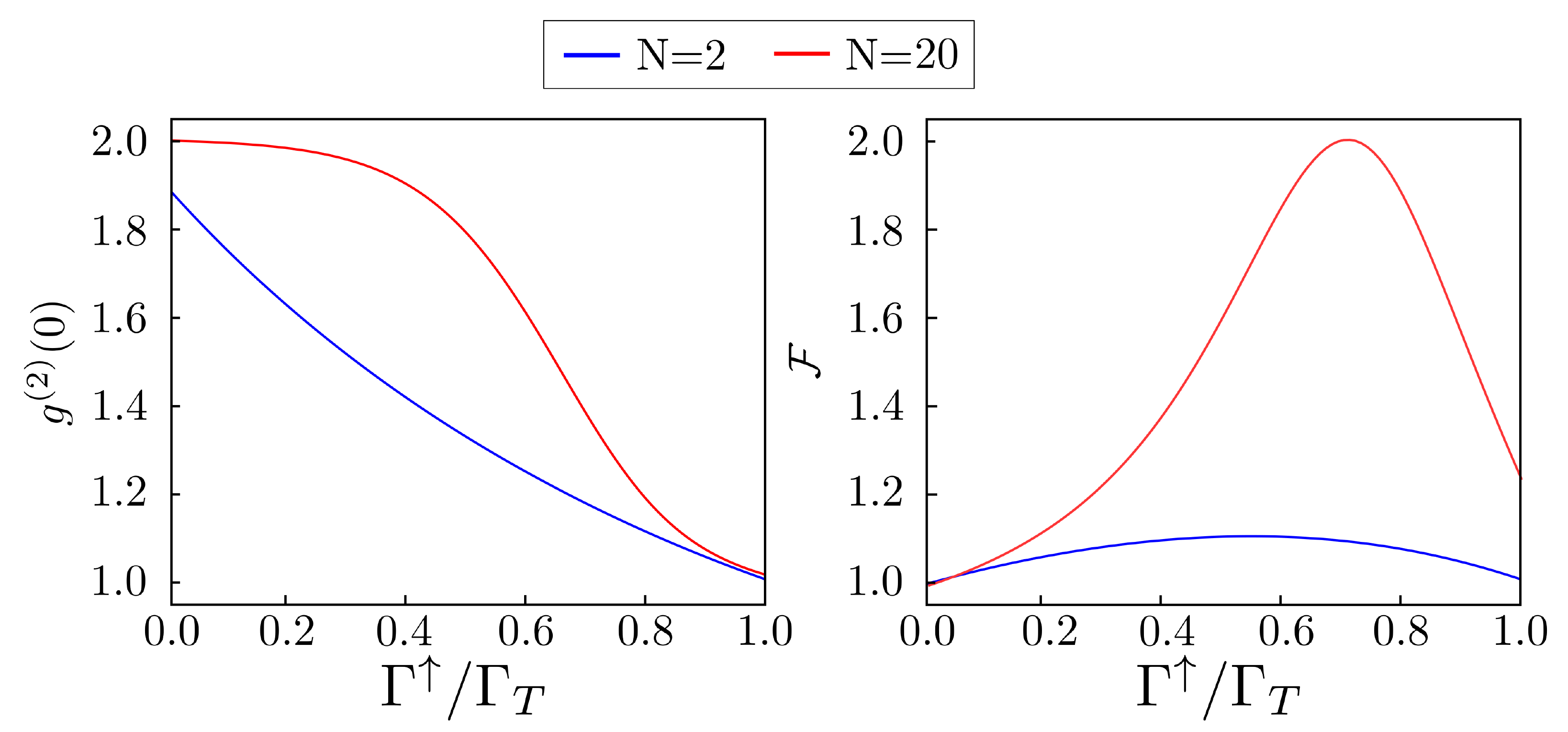}
    \abovecaptionskip=0pt
    \caption{Second order coherence  $g^{(2)}(0)$ and Fano factor $\mathcal{F}$ of the steady state photon distribution  for N=2 and N=20, 2-level systems. Excluding those labelled, the parameters are $\omega_{e}=\Gamma_{T}$, $g=0.9\Gamma_{T}$, and $\kappa = 0.8\Gamma_{T}$.}
    \label{fig:2LSFanog2}
  \end{minipage}}
\end{figure} 

Further in Fig.~\ref{fig:2LSFanog2} we show the second order coherence and Fano factor of the steady state photon distribution as defined in Eqs.~\eqref{eq:g2}--\eqref{eq:fano} of the main text for this 2-level model.  We see, much more clearly than in the 3-level model, that when $N$ is large there saturation of $g^{(2)}\simeq 2$ in the normal state which crosses over to $g^{(2)}\simeq 1$ in the lasing state. The accompanying Fano factor has a significant peak around the location of the transition.

\section{3-Level System: Additional Mean-Field Results}\label{app:3MFA}

\begin{figure}
\captionsetup{width=.9\linewidth}
  \lineskip=-\fboxrule
  \fbox{\begin{minipage}{\dimexpr \textwidth-2\fboxsep-2\fboxrule}
    \centering
    \includegraphics[width=12.5cm, angle=0]{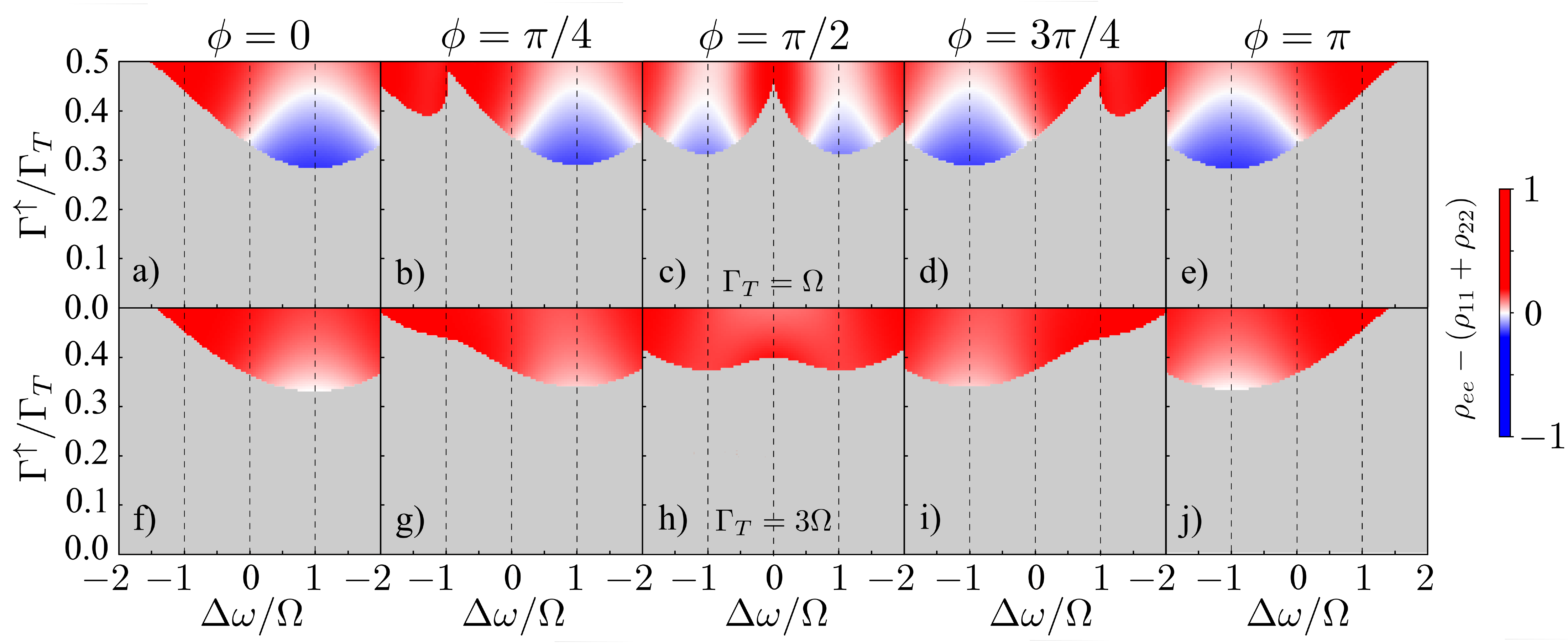}
    \abovecaptionskip=0pt
    \caption{Phase transition at different values of $\phi$.  In the white region the normal state is stable, while in the coloured regions a lasing instability occurs and the Jacobian matrix has a pair of eigenvalues with positive real part. The degree of inversion, and therefore lasing without inversion, is indicated by the colour gradient between blue and red. Excluding those stated, the parameters are:  $g=0.9\Omega$ and $\kappa = 0.8\Omega$.}
    \label{fig:PhiPhaseScan1}
  \end{minipage}}
\end{figure}

In Figure~\ref{fig:PhiPhaseScan1} we show a more complete set of results to go alongside those in Figure~\ref{fig:PhiPhaseScan} of the main text.
We see from the phase diagram that changing the phase by $\pi$ switches the $\ket{+}$ and $\ket{-}$ states and so  inverts the sign of the detuning.
The effects of changing the total atomic decay rate are also shown; we see that increasing $\Gamma_T$ uniformly increases the threshold pumping rate as the system needs to be more strongly driven to overcome the extra losses.
The qualitative features of the phase and detuning dependence are unaffected.
 
\begin{figure}
\captionsetup{width=.9\linewidth}
  \lineskip=-\fboxrule
  \fbox{\begin{minipage}{\dimexpr \textwidth-2\fboxsep-2\fboxrule}
    \centering
    \includegraphics[height=8.0 cm, angle=0]{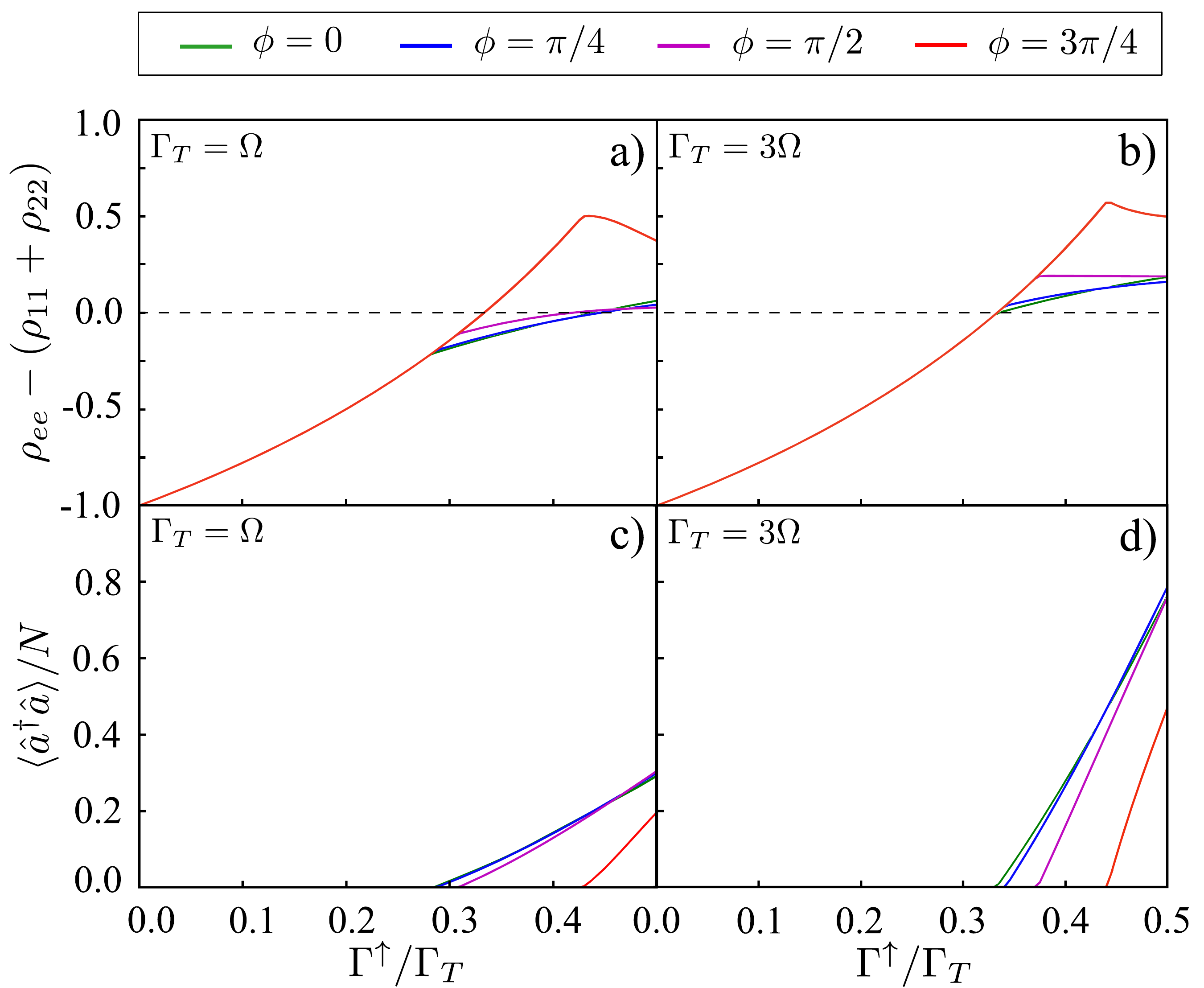}
    \abovecaptionskip=0pt
    \caption{Population inversion and cavity population at different values of  $\phi$ with a fixed detuning $\Delta \omega=\Omega$. Excluding those stated, the parameters are: $g=0.9\Omega$, and $\kappa = 0.8\Omega$.}
    \label{fig:OmInversion}
  \end{minipage}}
\end{figure}

We demonstrated in Figure~\ref{fig:3LSAtomPop} of the main text how the atomic population inversion and cavity occupation change as the system is driven through the lasing threshold for various microwave drive phases $\phi$.
These results were all obtained by optimising the detuning and tuning the system exactly to the correct resonance condition. Here, in Fig.~\ref{fig:OmInversion}, instead we show similar results, but instead keep the detuning fixed to $\Delta\omega=\Omega$. This again quantitatively changes the results but the behaviour is qualitatively the same as seen in the main text.

\begin{figure}
\captionsetup{width=.9\linewidth}
  \lineskip=-\fboxrule
  \fbox{\begin{minipage}{\dimexpr \textwidth-2\fboxsep-2\fboxrule}
    \centering
    \includegraphics[height=4.5 cm, angle=0]{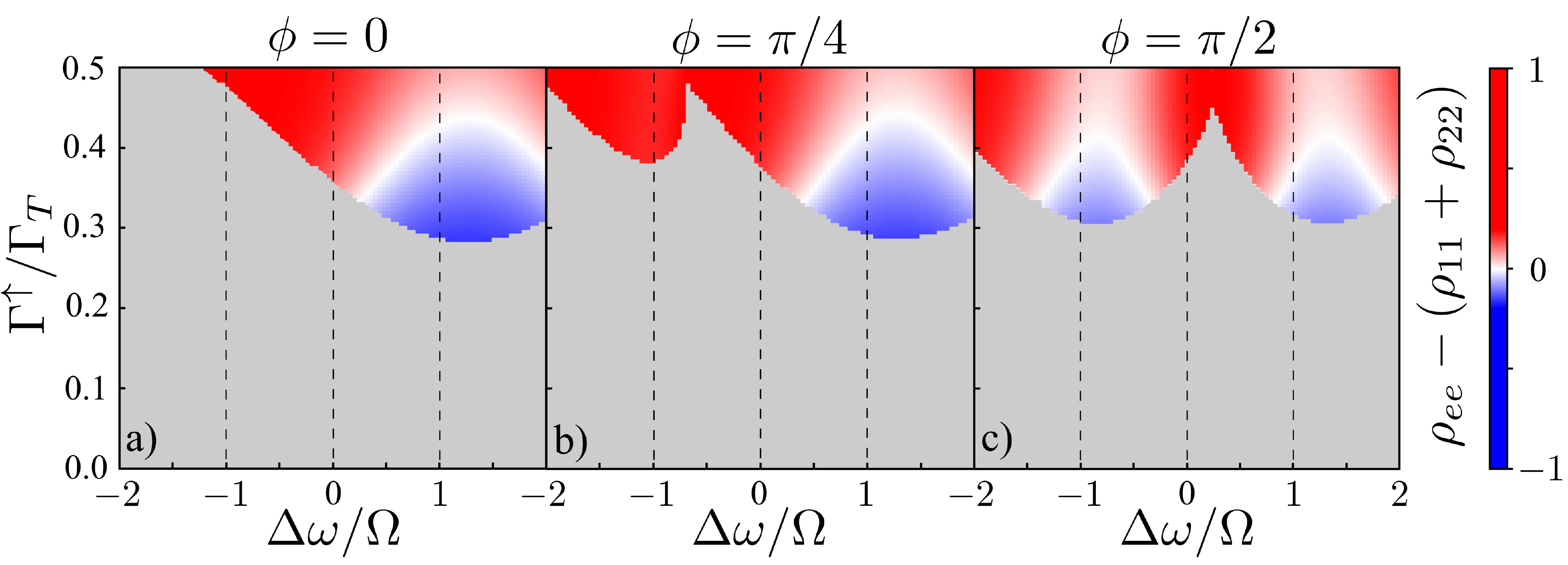}
    \abovecaptionskip=0pt
    \caption{Mean-field phase diagram showing the lasing regions at different values of $\phi$  with non-degenerate  lower levels $\omega_{e} -\omega_{1} =\Omega/2$ and  $\omega_{e} -\omega_{2} =\Omega$. In the white region the normal state is stable, while in the coloured regions a lasing instability occurs and the Jacobian matrix has a pair of eigenvalues with positive real part. In this region the degree of inversion, and therefore lasing without inversion, is indicated by the colour gradient. Excluding those stated, the parameters are $g=0.9\Omega$, $\kappa=0.8\Omega$, and $\Gamma_T = \Omega$.}
    \label{fig:PhiPhaseScanapp}
  \end{minipage}}
\end{figure}

\begin{figure}
\captionsetup{width=.9\linewidth}
  \lineskip=-\fboxrule
  \fbox{\begin{minipage}{\dimexpr \textwidth-2\fboxsep-2\fboxrule}
    \centering
    \includegraphics[height=9.0 cm, angle=0]{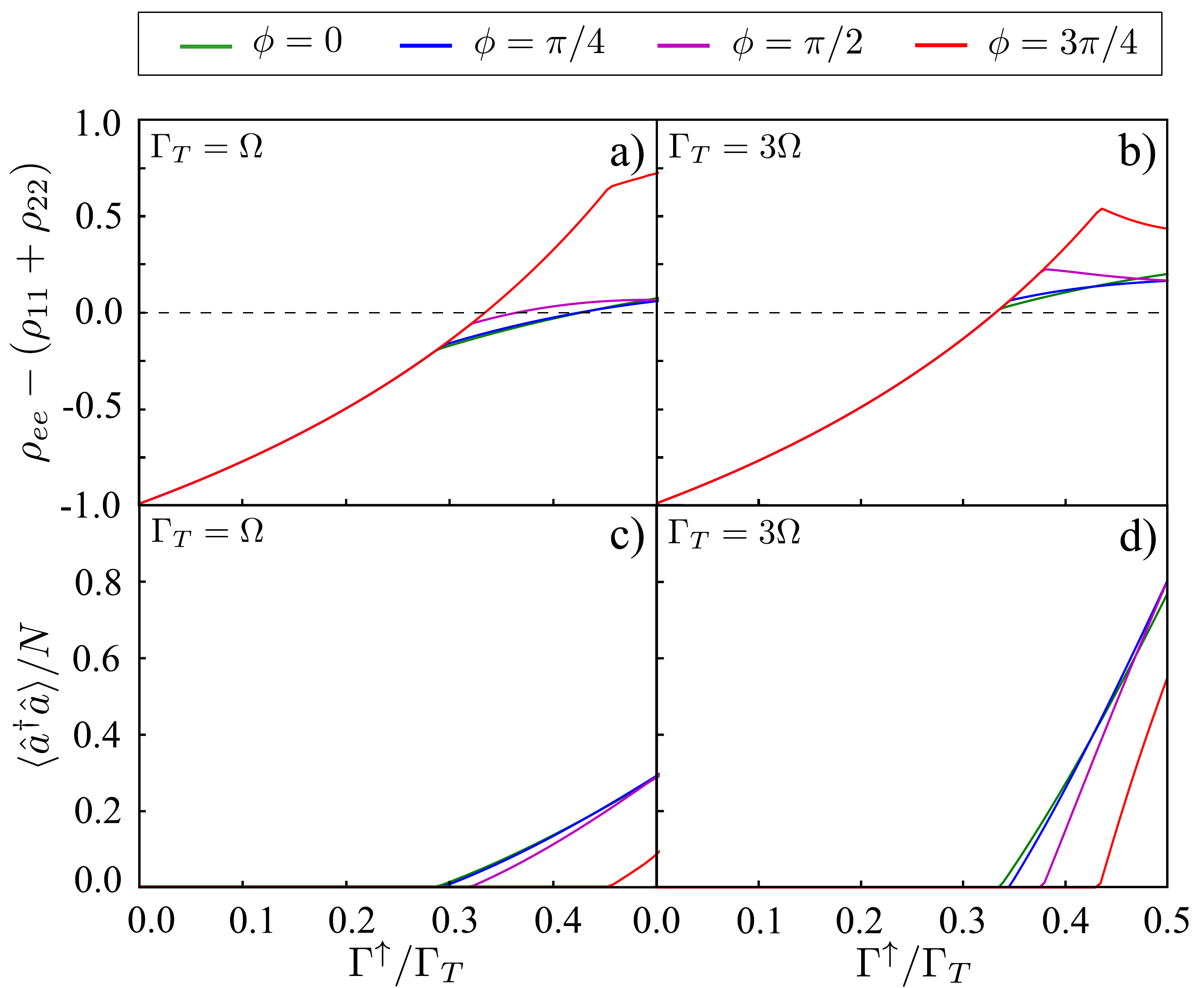}
    \abovecaptionskip=0pt
    \caption{Population inversion and cavity population at different values of $\phi$ with non-degenerate lower levels $\omega_{1} -\omega_{2} =\Omega/2$. Excluding those stated, the parameters are: $g=0.9\Omega$, and $\kappa = 0.8\Omega$.}
    \label{fig:OmInversiondetuning}
  \end{minipage}}
\end{figure}

We have thus far assumed that the two lower levels are degenerate. We now lift this assumption in order to show that the results presented in the main text are robust to the inclusion of this extra parameter. To do this we show the mean-field phase diagram  in Fig.~\ref{fig:PhiPhaseScanapp} and the steady state inversion and cavity population in Fig.~\ref{fig:OmInversiondetuning} both at finite detuning. We see very similar qualitative behaviour to that for the case where the two-lower levels are degenerate.

\section{3-Level System: Cumulant Equations}

For completeness, in this section we give the full set of second order cumulant equations which describe the 3-level system model. The procedure behind generating these is the same as that presented in~\ref{app:2QM}.

The first set of equations are for the first moments of the atomic operators. We again only include those terms which do not break the symmetry of the model. These are just the same as the mean-field equations presented in the main text except here we do not break the second order correlations which appear
\begin{gather}
    \braket{\dot{\rho}_{ee}} = -\frac{g}{\sqrt{N}}\left[i(\braket{a\rho_{e1}} + \braket{a\rho_{e2}}) + \text{c.c.} \right] + \Gamma^{\uparrow}(\braket{\rho_{11}}+\braket{\rho_{22}}) - 2\Gamma^{\downarrow}\braket{\rho_{ee}}, \\
    \braket{\dot{\rho_{11}}} = \left[i\left(\frac{g}{\sqrt{N}}\braket{a\rho_{e1}} - \Omega \text{e}^{-i\phi}\braket{\rho_{12}}\right) + \text{c.c.}\right]    - \Gamma^{\uparrow}\braket{\rho_{11}} + \Gamma^{\downarrow}\braket{\rho_{ee}}, \\
    \braket{\dot{\rho_{22}}} = \left[i\left(\frac{g}{\sqrt{N}}\braket{a\rho_{e2}} + \Omega \text{e}^{-i\phi}\braket{\rho_{12}}\right) + \text{c.c.}\right] - \Gamma^{\uparrow}\braket{\rho_{22}} + \Gamma^{\downarrow}\braket{\rho_{ee}}, \\
    \braket{\dot{\rho_{12}}} = i\left[\frac{g}{\sqrt{N}}\left(\braket{a\rho_{e2}} - \braket{a\rho_{e1}}^{*}\right) - \Omega \text{e}^{i\phi}(\braket{\rho_{11}} - \braket{\rho_{22}})\right]
    - \Gamma^{\uparrow}\braket{\rho_{12}}.
\end{gather}
The relevant second order terms which describe correlations between different atoms are given by
\begin{multline}
    \braket{ \dot{\rho_{1e} \rho_{e1}}} = \left( i\left[\frac{g}{\sqrt{N}} \braket{a\rho_{e1}} (\braket{\rho_{ee}} - \braket{\rho_{11}} - \braket{\rho_{12}}) - \Omega \text{e}^{-i\phi}\braket{ \rho_{e2} \rho_{1e}}\right] + \text{c.c.}\right) \\
    - \Gamma_{\phi}\braket{ \rho_{1e} \rho_{e1}},
\end{multline}
\begin{multline}
    \braket{\dot{\rho_{2e} \rho_{e2}}} = \left(i\left[ \frac{g}{\sqrt{N}} \braket{a\rho_{e2}} (\braket{\rho_{ee}} - \braket{\rho_{22}} - \braket{\rho_{12}}^{*}) + \Omega \text{e}^{-i\phi}\braket{\rho_{e2} \rho_{1e}}\right] + \text{c.c.}\right)  \\
    - \Gamma_{\phi}\braket{ \rho_{2e} \rho_{e2}},
\end{multline}
\begin{multline}
    \braket{ \dot{\rho_{e2} \rho_{1e}}} = i \frac{g}{\sqrt{N}} \left[\braket{a\rho_{e2}} \left(\braket{\rho_{ee}} - \braket{\rho_{11}} - \rho_{12}\right)  - \braket{a\rho_{e1}}^{*} \left(\braket{\rho_{ee}} - \braket{\rho_{22}} - \rho_{12}\right)\right]  \\
    - i\Omega \text{e}^{i\phi}\left(\braket{ \rho_{1e} \rho_{e1}} - \braket{ \rho_{2e} \rho_{e2}}\right) - \Gamma_{\phi}\braket{ \rho_{e2} \rho_{1e}},
\end{multline}
where $\Gamma_{\phi} = (2\Gamma_{\downarrow} + \Gamma_{\uparrow})/2$. Again here the two operators which appear act on different atoms and we have used the permutation symmetry of the model to realise that for each pair these must be identical. We have also broken the third order correlations which instead appear as products of first and second moments following the same procedure described in~\ref{app:2QM}.

There is then a set of terms which describe the correlations between the atoms and the field. These are given by
\begin{multline}
    \braket{\dot{a \rho_{e1}}} = -i\frac{g}{\sqrt{N}}\left[(N-1)\left(\braket{ \rho_{1e} \rho_{e1}} + \braket{ \rho_{e2} \rho_{1e}}^{*}\right) + \braket{\rho_{ee}}\right]  - i\Omega \text{e}^{-i\phi}\braket{a \rho_{e2}} \\
    -i\frac{g}{\sqrt{N}}\braket{a^{\dag}a}(\braket{\rho_{ee}} - \braket{\rho_{11}} - \braket{\rho_{12}}^{*}) -\left[i\left(\nu - \omega_{e}\right) + \frac{1}{2}\left(\kappa + \Gamma_{\phi}\right)\right]\braket{a \rho_{e1}},
\end{multline}
\begin{multline}
    \braket{\dot{a \rho_{e2}}} = -i\frac{g}{\sqrt{N}}\left[(N-1)\left(\braket{ \rho_{2e} \rho_{e2}} + \braket{ \rho_{e2} \rho_{1e}}\right) + \braket{\rho_{ee}}\right]  - i\Omega \text{e}^{i\phi}\braket{a \rho_{e1}}\\
    -i\frac{g}{\sqrt{N}}\braket{a^{\dag}a}\left(\braket{\rho_{ee}} - \braket{\rho_{22}} - \braket{\rho_{12}}\right) -\left[i\left(\nu - \omega_{e}\right) + \frac{1}{2}\left(\kappa + \Gamma_{\phi}\right)\right]\braket{a \rho_{e2}}. 
\end{multline}
Finally, the only second moment of the field required is the photon number operator which has the equations of motion
\begin{equation}
    \braket{\dot{a^{\dag}a}} = - \kappa \braket{a^{\dag}a} + g\sqrt{N}\left[ i\left(\braket{a\rho_{e1}} + \braket{a\rho_{e2}} \right)+ \text{c.c.}\right].
\end{equation}
Together the solutions to this set of equations can be used to calculate the behaviour of any quantity of interest.

\end{document}